\documentclass[prb,amsmath,twocolumn,reprint,letterpaper]{revtex4-2}
\usepackage{graphicx}
\usepackage{amssymb}
\begin{document}

\title{Benchmarking the accuracy of superconducting pair-pair
  correlations within Constrained Path Quantum Monte Carlo}
\author{Jodie Roberts$^1$}
\author{Beau A.~Thompson$^2$} 
\author{R.~Torsten Clay$^2$} \email{r.t.clay@msstate.edu}
\affiliation{$^1$Department of
  Engineering Science and Mechanics, Pennsylvania State University, University Park, PA 16802}
\affiliation{$^2$Department of Physics \& Astronomy, and Applied
    Research Collaboratory
  Center for Computational Sciences, Mississippi State University,
  Mississippi State, MS 39762} 
\date{today}
 \begin{abstract}
 Ground state properties of the Hubbard model are of fundamental
 importance to understand the mechanism of unconventional
 superconductivity in the high-T$_c$ cuprates and other materials.
 One of the most powerful numerical methods for strongly interacting
 models is quantum Monte Carlo, which however faces a fundamental
 limitation, the Fermion sign problem.  The sign problem can be
 mitigated using approximate methods such as Constrained Path Monte
 Carlo, where the sign of the overlap with a trial
   wavefunction is constrained to remain positive. Beyond the
   constraint, additional approximations must be made in order to
   measure observables for operators that do not commute with the
   Hamiltonian. We examine critically the most commonly used
 approximation, back propagation, as well as a recently proposed
 constraint release measurement technique. In comparisons with a
 variety of systems that can be solved numerically exactly by other
 methods, we find that back propagation tends to underestimate
 long-range superconducting pair-pair correlations. 
   Except in special cases such as one dimension, back propagation
   introduces an uncontrolled approximation even when the exact trial
   wavefunction is used. The constraint release technique can provide
 accurate results, with the disadvantages that it much more
 computationally expensive and reintroduces the sign problem.
 \end{abstract}
\maketitle

\section{Introduction}

The Hubbard model has played a central role in the theory of strongly
correlated materials \cite{Arovas22a}. It has been widely used as a
base theoretical model for unconventional superconductors, including
the high-T$_{\rm c}$ cuprates \cite{Keimer15a} and the organic
charge-transfer solid superconductors \cite{Clay19a}.  Numerical
many-body methods have been essential to understanding its properties
\cite{Qin22a}, and in turn, the study of the Hubbard model has driven
the development of these methods.  Among many-body methods, quantum
Monte Carlo (QMC) is one of the most used \cite{Gubernatis16a}.  The
primary limiting factor for QMC studies of the Hubbard model is the
Fermion sign problem, which leads to an exponential loss of precision
with increasing lattice size, inverse temperature, and interaction
strength \cite{White89a,Imada89a}.

Constrained Path Monte Carlo (CPMC) is a QMC method that mitigates the
sign problem \cite{Zhang95a,Zhang97a}.  Based on the auxiliary field
determinant Monte Carlo method \cite{Gubernatis16a}, CPMC provides an
efficient although approximate solution to the sign problem by using a
trial wavefunction. While the error introduced by the trial function
is unknown, one can in principal improve the results by using a more
complex trial function. Ground state energies from CPMC compare well
against known exact values and other methods \cite{Zhang97a,Qin16a}.
Because of this, CPMC has been widely used to study superconducting
pairing in Hubbard models
\cite{Zhang97a,Zhang97b,Guerrero98a,Guerrero99a,Guerrero00a,Huang01a,
  Gomes16a,DeSilva16a,Clay19b,Qin20a,Xu24a,Chen25a}.  The earliest
CPMC calculations found no evidence for superconductivity (SC) in the
doped Hubbard model on a square lattice \cite{Zhang97b,Guerrero99a},
confirmed by later calculations \cite{Qin20a}. Beyond the square
lattice, interactions were found to enhance SC near quarter or
three-quarter filling in anisotropic triangular \cite{Gomes16a} and
other frustrated lattices \cite{DeSilva16a,Clay19b}.  Recent CPMC
calculations using with boundary-condition averaging have suggested
that SC is present in the thermodynamic limit in the lightly doped
Hubbard model with next-nearest neighbor hopping \cite{Xu24a}.

A question of central importance is how accurate are methods like
CPMC. Most tests of accuracy have focused on comparisons of the ground
state energy between CPMC, other methods, and exact results. While
this is useful, correlated many-body states are often separated by
very small energy differences that are difficult to resolve. Energy
comparisons are also complicated by the fact that CPMC is not a
variational method, with the energy calculated by CPMC
sometimes being below the true ground state energy
\cite{Carlson99a}.  In this work we examine critically the accuracy of
CPMC in calculating correlation functions, specifically the equal-time
superconducting pair-pair correlation.  In order to calculate
correlation functions of operators that do not commute with the
Hamiltonian, an additional approximation, back propagation
\cite{Zhang97a,Motta17a}, must be used. We compare this with the
recently proposed constraint release method \cite{Xiao23a}.  We will
show that over a wide variety of lattices and parameters, CPMC with
back propagation tends to underestimate the magnitude of
long-range superconducting pair-pair correlations. Furthermore,
   back propagation does not in general produce
  statistically  exact correlations even when the exact trial
  function is available.

Below in Section \ref{method} we define the model and pairing
observables we measure, and briefly review the CPMC method and the
back propagation and constraint release techniques. In Section
\ref{results} we present computational results for several systems,
including one dimension, two-leg ladders, the square lattice, and the
anisotropic triangular lattice.  We also demonstrate a technique for
generating high accuracy trial wavefunctions which can in principle be
systematically improved using the Path Integral Renormalization Group
(PIRG) method \cite{Imada00a,Kashima01a,Mizusaki04a}. For
  small systems, PIRG wavefunctions also allow us to investigate the
  properties of CPMC when the {\it exact} trial function is used.
Section \ref{conclusions} summarizes our findings. Appendices
\ref{appendixa} and \ref{appendixb} discuss the use of
  CPMC in one dimensional systems and present additional test data.

\section{Model and Methods}
\label{method}

\subsection{Hubbard model and observables}

We consider the Hubbard model with Hamiltonian
\begin{eqnarray}
  \hat{H} &=& - \sum_{\langle ij\rangle,\sigma}t_{ij}(c^\dagger_{i,\sigma}c_{j,\sigma} + H.c.)
  + U\sum_i n_{i,\uparrow}n_{i,\downarrow}.  \label{ham}
\end{eqnarray}
In Eq.~\ref{ham}, $c^\dagger_{i,\sigma}$ creates an electron of spin
$\sigma$ on site $i$ and
$n_{i,\sigma}=c^\dagger_{i,\sigma}c_{i,\sigma}$.  $U$ is the onsite
Coulomb interaction.  In the first sum, $\langle ij\rangle$ denotes
all bonds of the lattice. While we will consider several possible
lattice structures, in each case we give energies in units of the
largest hopping integral among the $t_{ij}$.

In addition to measurements of the ground state energy, we consider
measurements of the equal-time superconducting pair-pair correlation.
We define the singlet pair creation operator as
\begin{equation}
\Delta^\dagger_i=\frac{1}{\sqrt{2}}(c^\dagger_{i_1,\uparrow}c^\dagger_{i_2,\downarrow}-
c^\dagger_{i_1,\downarrow}c^\dagger_{i_2,\uparrow})\label{Delta},
\end{equation}
which creates a singlet pair on sites $i_1$ and $i_2$. The pair-pair
correlation is then
\begin{equation}
  P(r\equiv|{\bf r}_i-{\bf r}_j|) = \frac{1}{2}\langle \Delta^\dagger_{{\bf r}_i} \Delta_{{\bf r}_j}
+  \Delta_{{\bf r}_i} \Delta_{{\bf r}_j}^\dagger \rangle.
  \label{pr1d}
\end{equation}
SC in a two-dimensional (2D) system requires long-range order in $P(r)$
at zero temperature,
with $P(r)$ reaching a finite value as $r\rightarrow\infty$.
On a  2D lattice the appropriate pair creation
operator is a superposition of four nearest-neighbor pairs,
\begin{equation}
  \Delta^\dagger_{{\bf r}}=\mathcal{N}\sum_{\nu=1}^4g_\nu(c^\dagger_{{\bf r},\uparrow}c^\dagger_{{\bf r}+{\bf r}(\nu),\downarrow} -  c^\dagger_{{\bf r},\downarrow}c^\dagger_{{\bf r}+{\bf r}(\nu),\uparrow}).
  \label{pr}
\end{equation}
In Eq.~\ref{pr}, $\mathcal{N}$ is a normalization factor and
$g_\nu$ a relative phase.  $g_\nu$ are all positive for
$s$-wave pairing, while for $d_{x^2-y^2}$ pairing $g_\nu$=$\{+1,-1,+1,-1\}$
corresponding to ${\bf r}(\nu)=\{\hat{x},\hat{y},-\hat{x},-\hat{y}\}$.
We further define the
 average long-range pair-pair correlation,
\begin{equation}
  \bar{P}=\frac{1}{\mathcal N_p} \sum_{r_i>2} P(r_i).
  \label{pbar}
\end{equation}
In $\bar{P}$ only correlations beyond a minimum distance 
of two lattice units are considered. The normalization $\mathcal N_p$ is
the number of correlations that satisfy $r_i>2$. Because pair-pair correlations
are much stronger at very short distances, $\bar{P}$ is a useful quantity
to evaluate the tendency towards superconductivity on finite
lattices \cite{Gomes16a}.

\subsection{Constrained Path Monte Carlo}

We only briefly review the CPMC method here, as further details are
available in the literature
\cite{Zhang97a,Gubernatis16a,Motta17a,Xiao23a}.  In CPMC the ground
state wavefunction is represented as a sum over random walkers,
\begin{equation}
|\Psi_0\rangle = \sum_i w_i |\phi_i\rangle,
\end{equation}
with each walker in the ensemble having a weight $w_i$ and Slater
determinant wavefunction $|\phi_i\rangle$. $|\phi_i\rangle$ actually
represents a direct product of independent spin up and spin down
determinants; for clarity we suppress the spin indices below.  CPMC is
a projector technique where the ground state wavefunction
$|\Psi_0\rangle$ is projected from an initial state $|\Psi_i\rangle$
in imaginary time,
\begin{equation} 
|\Psi_0\rangle = \lim_{\beta\rightarrow\infty} e^{-\beta(\hat{H}-E_0)}|\Psi_i\rangle,
\end{equation}
with $E_0$ a trial energy. The exponential of $\hat{H}$ is  discretized 
typically with a second-order Trotter approximation,
\begin{equation}
  e^{-\Delta\tau\hat{H}}\approx e^{-\Delta\tau/2\hat{K}}e^{-\Delta\tau\hat{U}}e^{-\Delta\tau/2\hat{K}},
  \label{trotter2nd}
\end{equation}
where $\hat{K}$ and $\hat{U}$ are the hopping and interaction parts of Eq.~\ref{ham}
respectively, and $\Delta\tau$ is the imaginary time step.
While the exponential of $\hat{K}$ can be calculated exactly,
the interaction term  is transformed using a Hubbard-Stratonovich
transformation,
\begin{equation}
e^{-\Delta\tau U n_{i\uparrow}n_{i\downarrow}}=e^{-\Delta\tau U(n_{i\uparrow}+n_{i\downarrow})/2}\sum_{x_i=\pm1}
\frac{1}{2}e^{\gamma x_i(n_{i\uparrow}-n_{i\downarrow})},
\label{hs}
\end{equation}  
with $\cosh\gamma=e^{\Delta\tau U/2}$. An open-ended random walk is
then performed on each random walker, where the value of the field
$x_i$ at each site is selected via Monte Carlo
  sampling. Straightforward Monte Carlo sampling of Eq.~\ref{hs}
  assigns an equal probability to the values of $x_i=\pm 1$,
  $P(x_i)=\frac{1}{2}$.  A central part of the CPMC algorithm is the
  use of importance sampling, with a modified probability $\tilde{P}$
  \cite{Zhang97a,Gubernatis16a}. Each random walker represented by the
  Slater determinant $|\phi\rangle$ is weighted by its overlap with a
  trial wavefunction $|\Psi_t\rangle$, $O_T(\phi)=\langle
  \Psi_t|\phi\rangle$.  The modified probability $\tilde{P}$ is then
    \begin{equation}
      \tilde{P}(x_i)=\frac{O_T(\phi^\prime)}{O_T(\phi)}P(x_i).
      \label{importance}
    \end{equation}
    In Eq.~\ref{importance}, $\phi$ ($\phi^\prime$) are the Slater determinants of the walker
    before (after) application
    of the field in Eq.~\ref{hs}.
  The projection in imaginary time is further
  subjected to the constraint $\langle \Psi_t|\phi_i\rangle > 0$, with
  random walkers removed from the ensemble if their overlap becomes
  negative.  The trial wavefunction $|\Psi_t\rangle$ thus plays a
  critical role in the method. In the noninteracting limit
    ($U=0$) the method becomes trivially exact.  We will assume that
  $|\Psi_i\rangle = |\Psi_t\rangle$ and that $|\Psi_t\rangle$ is a sum
  of individual Slater determinants,
  \begin{equation}
  |\Psi_t\rangle = \sum_k^{N_t} c_k |\phi^t_k\rangle.
  \end{equation}
      The CPMC method computationally scales as $\sim N^3$ based on
      the linear algebra of propagating walkers. In the case of a
      multi-determinant trial function this must be multiplied by
      $N_t$.

  \subsection{Mixed estimator}
\label{mixedsection}

 Several techniques are used to measure observables in CPMC. The
 fundamental problem is that to measure the expectation value of a
 general operator in the ground state wavefunction, $\langle \hat{\mathcal{O}}\rangle
 = \langle \Psi_0 | \hat{\mathcal{O}} | \Psi_0 \rangle$, {\it two} independent
 stochastic samples of $|\Psi_0\rangle$ would be required
 \cite{Carlson99a}, which is not feasible in practice. In the
 case where $\hat{\mathcal{O}}$ commutes with $\hat{H}$, the mixed
 estimator can be used \cite{Zhang97a,Motta17a},
 \begin{equation}
  \langle \mathcal{O}_{mix}\rangle = \frac{1}{\sum_i w_i} \sum_i w_i
  \frac{\langle \Psi_t|\hat{\mathcal O}|\phi_i\rangle}{\langle \Psi_t|\phi_i\rangle}.
   \label{mixed}
 \end{equation}
  Such measurements are taken periodically 
 in imaginary time and binned in order to decrease correlations.
 It has been observed that even with simple choices for $|\Psi_t\rangle$, such
 as a single-determinant non-interacting wavefunction, CPMC using the mixed
 estimator can calculate energies with errors of a few percent \cite{Zhang97a}.
 One important limitation of the mixed energy estimate is that it
 is not variational \cite{Carlson99a}.

 \subsection{Back propagation}

 In practice, Eq.~\ref{mixed} can only be used for $\hat{\mathcal
   O}=\hat{H}$; for other operators it produces very inaccurate
 results. For all other expectation values, the back propagation (BP)
 scheme is used \cite{Zhang97a,Motta17a}. In BP, $\langle \Psi_t|$ in
 Eq.~\ref{mixed} is modified with a string of propagators 
   \cite{Zhang97a,Motta17a},
 \begin{equation}
    \langle \mathcal{O}_{BP}\rangle = \frac{1}{\sum_i w_i}
   \sum_i w_i \frac{\langle \Psi_t|e^{-\beta \hat{H}}\hat{\mathcal O}|\phi^0_i\rangle}
   {\langle \Psi_t|e^{-\beta \hat{H}}|\phi^0_i\rangle}.
   \label{bp}
\end{equation}   
 Here $\beta=N_{\rm BP}\Delta\tau$ where $N_{\rm BP}$ is the number of
 time steps over which BP is performed. BP is implemented on top of
 the open-ended random walk of the CPMC method
 \cite{Zhang97a,Motta17a}.  At the start of a BP measurement, the
 Slater determinants for each walker are saved as
 $|\phi_i^0\rangle$. Then over the next $N_{\rm BP}$ Monte Carlo steps
 the Hubbard-Stratonovich fields $x_i$ are saved. To perform the
 measurement, the individual propagators making up $e^{-\beta\hat{H}}$
 are applied to $|\Psi_t\rangle$ in the reverse order using the saved
 $x_i$. The weights $w_i$ in Eq.~\ref{bp} are the current weights of
 each walker when each measurement is taken. BP introduces unknown
 errors into the measurements, because the overlap constraint imposed
 on walkers is not necessarily valid in the reverse direction
 \cite{Motta17a}. $\beta$ must be large enough to ensure $\langle
 \Psi_t|e^{-\beta\hat{H}}$ is distinct from $\langle
 \Psi_t|$. One way to quantify this is to calculate the
   overlap $\langle\Psi_t|e^{-\beta\hat{H}}|\Psi_t\rangle$ during back
   propagation (see for example Fig.~\ref{bp-overlap} in Appendix
   \ref{appendixb}).  Larger  $\beta$ however decreases the
 number of measurements and increases statistical errors.

 \subsection{Constraint release}

Constraint release (CR) is a third measurement estimator that can in principal correct
for the systematic error introduced by the constrained path approximation.
We use the CR algorithm described in Reference \cite{Xiao23a}. The CR algorithm
is  based on the zero temperature projector quantum Monte Carlo
 \cite{Gubernatis16a}, where expectation values are evaluated as
 \begin{equation}
   \langle \hat{\mathcal O} \rangle \approx \frac{\langle \Psi_L|e^{-(\beta-\tau)\hat{H}}\hat{\mathcal O}
     e^{-\tau\hat{H}}|\Psi_R\rangle}{\langle \Psi_L|e^{-(\beta-\tau)\hat{H}}e^{-\tau\hat{H}}|\Psi_R\rangle}.
   \label{zerot}
 \end{equation}   
 $|\Psi_L\rangle$ and $|\Psi_R\rangle$ are trial wavefunctions and
 conventionally $|\Psi_L\rangle=|\Psi_R\rangle$.  $\tau$ controls the
 length of projection in imaginary time applied to $|\Psi_L\rangle$
 and $|\Psi_R\rangle$, with $\tau=\beta/2$ when the same trial
 wavefunction is used for both $|\Psi_L\rangle$ and $|\Psi_R\rangle$.
 As with CPMC, $e^{-(\beta-\tau)\hat{H}}$ and $e^{-\tau\hat{H}}$ are
 broken into slices of imaginary time $\Delta\tau$ and Eq.~\ref{hs} is
 used to replace the interactions with Hubbard-Stratonovich fields.
 Eq.~\ref{zerot} is estimated using Monte Carlo sampling,
\begin{equation}
  \langle \hat{\mathcal O} \rangle \approx \frac{\sum_{\bf x} \langle s({\bf x}) \hat{\mathcal O}({\bf x}) \rangle}{\sum_{\bf x}
    \langle s({\bf x}) \rangle}.
\label{mcsamp}
\end{equation}   
In Eq.~\ref{mcsamp}, {\bf x} represents the configuration of
Hubbard-Stratonovich fields in Eq.~\ref{zerot} on all lattice sites
and imaginary time slices. The Monte Carlo weight of {\bf x}, $W({\bf
  x})$, is given by the determinant of the matrices in the denominator
of Eq.~\ref{zerot} \cite{Gubernatis16a}.  In general, the weight
$W({\bf x})$ can become negative.
Monte Carlo updates proceed through ``sweeps'' of the
  lattice, where  each Hubbard-Stratonovich field is sequentially sampled.
The Monte Carlo sampling in
Eq.~\ref{mcsamp} is done with respect to $|W({\bf x})|$, with $s({\bf
  x})$ the sign of $W({\bf x})$. 
While the zero temperature projector
method in principle gives exact ground-state values for large enough
$\beta$ (assuming the trial functions are not orthogonal to
$|\Psi_0\rangle$), it suffers from the Fermion sign problem where
$\langle s({\bf x})\rangle$ becomes small, requiring an increasingly
large number of Monte Carlo samples to evaluate Eq.~\ref{mcsamp}. The
sign problem becomes intractable with larger $\beta$, increasing
system size, and stronger interactions.

One way to mitigate the sign problem in Eq.~\ref{zerot} is to improve
the trial wavefunctions $|\Psi_L\rangle$ and $|\Psi_R\rangle$, so that
a smaller imaginary projection time $\beta$ is required to project out
the ground state. This is the basis for the CR
technique for CPMC \cite{Xiao23a}. In this method,
$|\Psi_L\rangle=|\Psi_t\rangle$ and $|\Psi_R\rangle=|\phi^0_i\rangle$,
where $|\phi^0_i\rangle$ is the saved Slater determinant of random
walker $i$ as in Eq.~\ref{bp}. Because $|\phi^0_i\rangle$ comes from the
CPMC ensemble of walkers representing the ground state $|\Psi_0\rangle$,
it is expected that $|\langle\Psi_0|\phi^0_i\rangle| > |\langle\Psi_0|\Psi_t\rangle|$ and the sign problem encountered in Eq.~\ref{zerot} will be
less severe than if $|\Psi_t\rangle$ were used for both
$|\Psi_L\rangle$ and $|\Psi_R\rangle$.
This also suggests that the optimum choice of $\tau$ is $\tau < \beta/2$,
which we will demonstrate in Section \ref{results}.
The CR
estimate we use is given by \cite{Xiao23a}
\begin{equation}
  \langle  \mathcal{O}_{CR}\rangle =
  \frac{1}{\sum_i w_i}
  \sum_i w_i \frac{\sum_{\bf x} s({\bf x})\frac{\langle\Psi_t| e^{-(\beta-\tau)\hat{H}}\hat{\mathcal O}
      e^{-\tau\hat{H}}|\phi_i^0 \rangle}{\langle\Psi_t| e^{-(\beta-\tau)\hat{H}}\hat
      e^{-\tau\hat{H}}|\phi_i^0 \rangle}}
      {\sum_{\bf x} \langle s({\bf x}) \rangle}.
\label{cr}
\end{equation}
In Eq.~\ref{cr}, the fields ${\bf x}$ are initialized to their value
in the original CPMC path, as in the BP technique.
In principle CR corrects for the unknown error in CPMC measurements
due to the approximate nature of the constraint, at the cost of
reintroducing the Fermion sign problem.
One disadvantage compared to BP is that CR introduces more
 parameters for each measurement:   $\tau$,  the
   number of Monte Carlo sweeps over the fields {\bf x}, and
   the number of sweeps to discard before recording data.
The principal disadvantage of
using CR in actual calculations is that it is far slower than BP,
because for every measurement, a separate Monte Carlo calculation must
be done to sample over the fields {\bf
  x}.
  Each sweep of the CR measurement scales similarly to  the
  computation required for a single BP measurement ($\sim N^3$).
  Therefore, CPMC-CR is slower than CPMC-BP  by a factor of
  approximately  the number of sweeps used in each CR measurement.
  In practice CPMC-CR is somewhat faster than this, because to reduce data
  correlation it is  not necessary to perform measurements after
  every sweep. For example, in our 4$\times$4 calculations (see
  Section \ref{square}) using 200
  CR sweeps and measuring every 10th sweep with $\beta=1$, CPMC-CR was about
  60 times slower than CPMC-BP.
The increased computational time can however be offset
by parallel computing, which is simple to implement for Monte Carlo
methods. An additional technique that can be used
  to speed up the CR measurement sweeps is the delayed update,
  which by grouping several updates together takes advantage of 
  memory caching on modern CPUs \cite{Du25a}.

\section{Results}
\label{results}

In this section we present comparisons of CPMC with  exact results
on  a variety of different lattices: one dimension,
two-leg ladders, and periodic 2D clusters both with and without
lattice frustration. We also investigate the effect
  of different choices for $|\Psi_t\rangle$, including the
  choice $|\Psi_t\rangle=|\Psi_0\rangle$.

 \subsection{One dimension}
\label{1d}
 
 In one dimension (1D) with only nearest-neighbor hopping
  it has been observed that the constrained path approximation is
  exact \cite{Clay99a,Carlson99a}. When compared with numerically
  exact Stochastic Series Expansion (SSE) QMC results, CPMC produces
  identical results within statistical error for the mixed estimator
  energy when using a free-electron trial function
  \cite{Clay99a}. This remains true when nearest-neighbor Coulomb
  interactions are considered in addition to $U$ \cite{Clay99a}.
  Specifically, the method appears to be exact for uniform 1D systems
  with odd (even) $N_\sigma$ in periodic (anti-periodic) boundary
  conditions 
  ($N_\sigma$ is the number of particles of a given spin).  As we
  discuss further in Appendix \ref{appendixa}, this is related to the
  positivity of the coefficients of $|\Psi_0\rangle$ in 1D, and one
  can show that the CPMC constraint is never violated when using the
  free-electron determinant as $|\Psi_i\rangle$.
  Unlike several other QMC methods, CPMC is not in general exact
  for 1D systems with open boundaries.
 \begin{figure}[tb]
   \begin{center}
     \resizebox{3.2in}{!}{\includegraphics{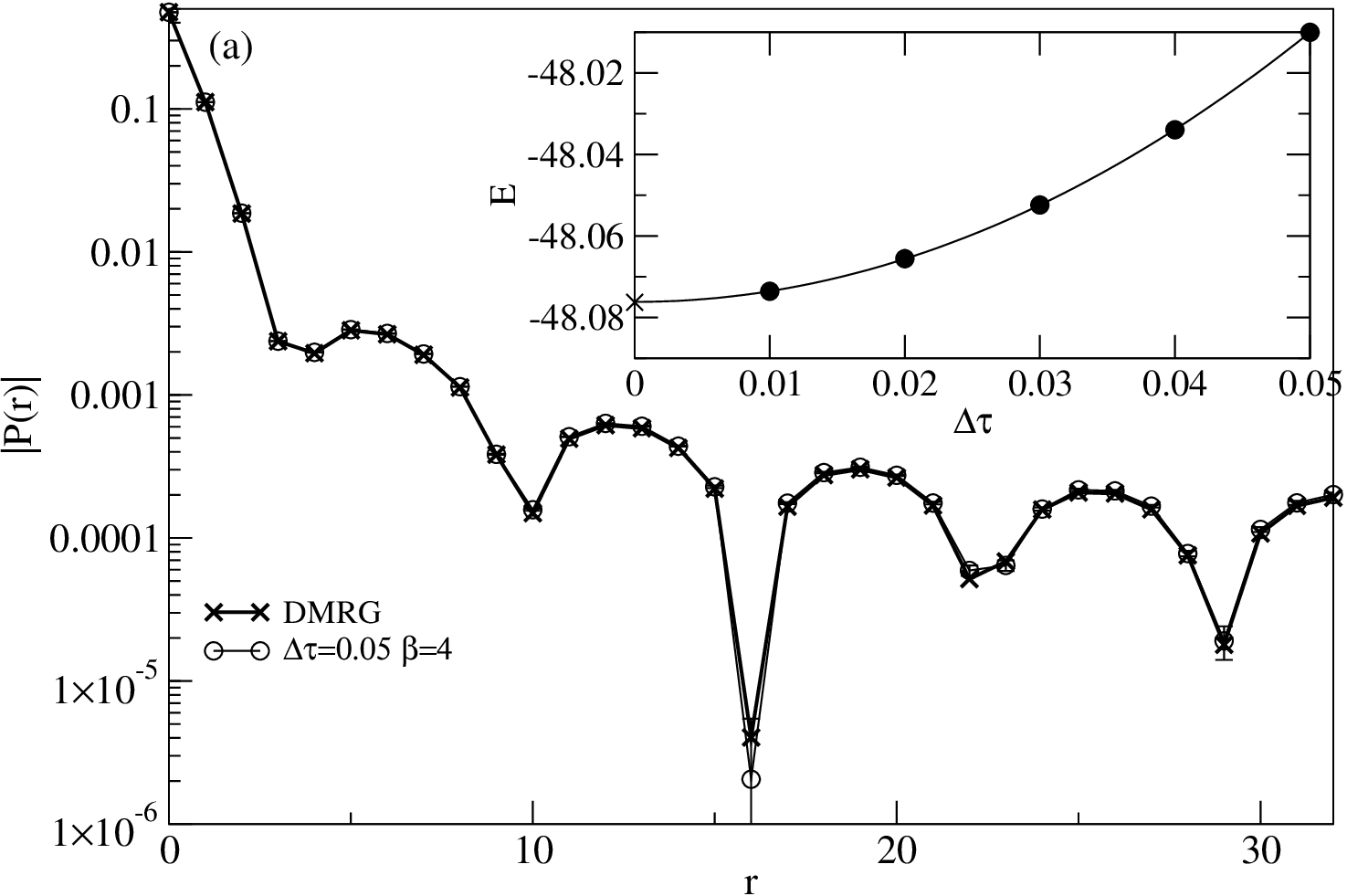}}
     \resizebox{3.2in}{!}{\includegraphics{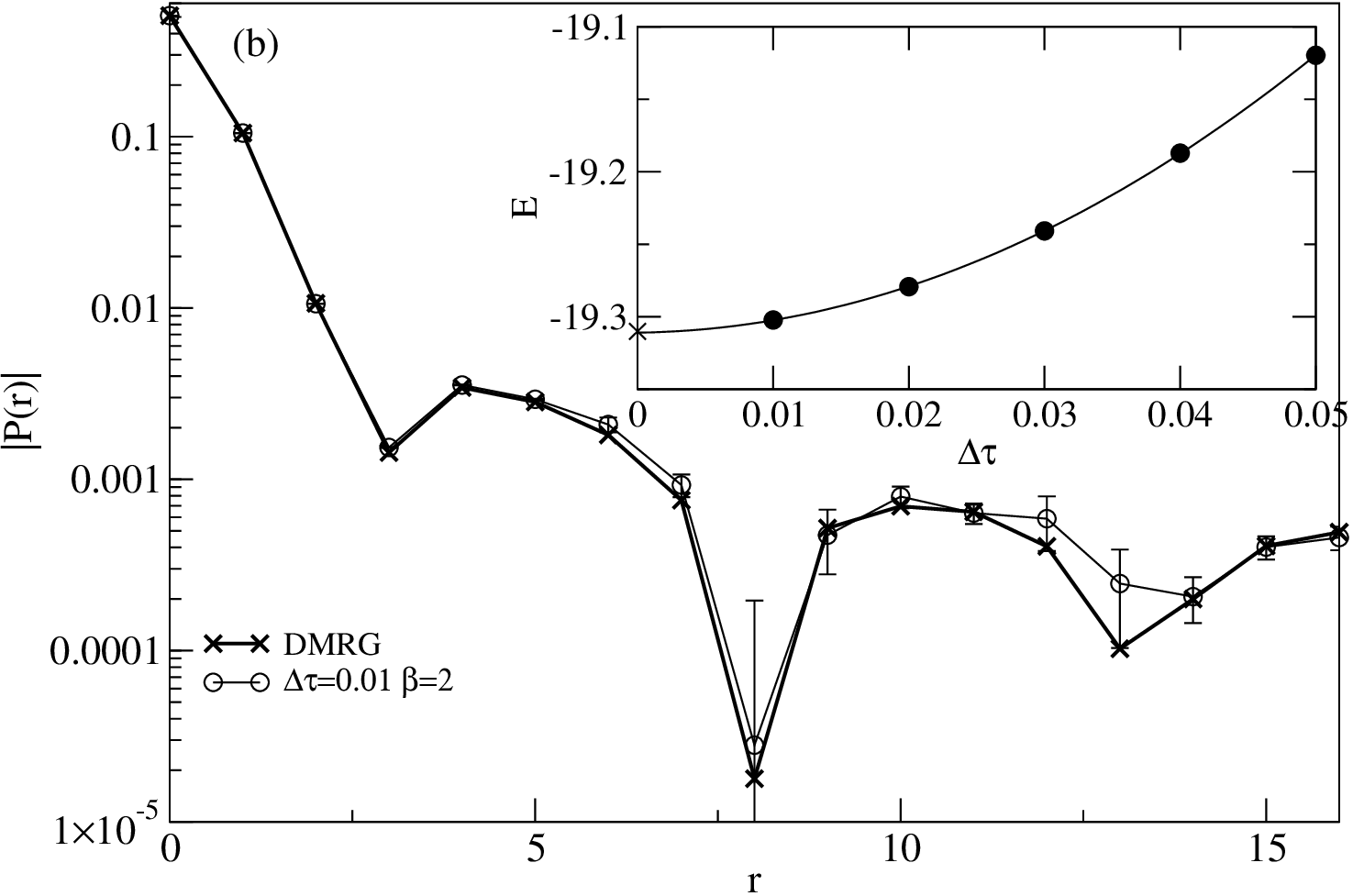}}
     \resizebox{3.2in}{!}{\includegraphics{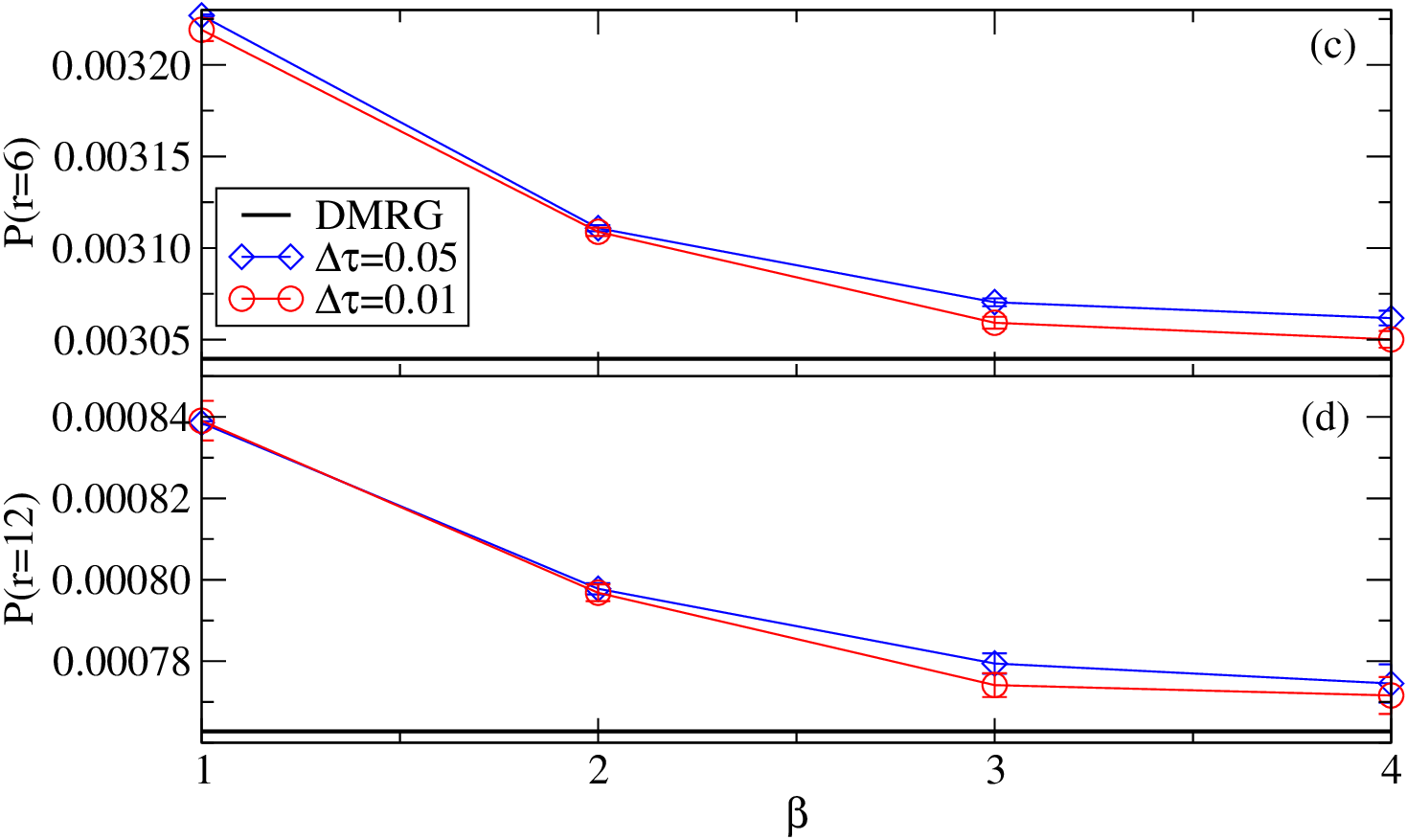}}
   \end{center}
   \caption{(color online) (a) Pair-pair correlation for nearest-neighbor
     pairs, $P(r)$, versus distance $r$ for a $N$=64 site one-dimensional
     lattice with periodic boundary conditions, $N_p$=54 particles, and
     $U$=4. The x's are exact results calculated using DMRG;
     circles are calculated using CPMC with BP.  The inset shows the
     convergence of the CPMC mixed energy estimator versus the
     imaginary time step $\Delta\tau$. The line is a quadratic fit to
     the CPMC data; the x at $\Delta\tau$=0 is the energy
     calculated by DMRG. (b) Same as (a), except $N$=32, $N_p$=26, and
     $U$=8. (c) and (d) Convergence of $P(r)$ as a function of $\Delta\tau$
     and $\beta$ for selected $r$ for $N$=32, $N_p$=26, and $U$=4.
      The heavy line is the DMRG result. \label{1ddata}}
 \end{figure}

In Fig.~\ref{1ddata}(a) we show CPMC data for a 64 site periodic
chain with $U=4$, using the $U=0$ wavefunction for $|\Psi_t\rangle$
and 8000 random walkers.  The total length of projection in imaginary
time was 1.5$\times$10$^5$.  We calculated numerically exact results
for the same system using the Density Matrix Renormalization Group
(DMRG) method and the ITensor library \cite{itensor}. As shown in the
inset of Fig.~\ref{1ddata}(a), the mixed energy is numerically exact when extrapolated to
remove Trotter error.  In a much more sensitive
test, we measure $P(r)$ for singlet nearest-neighbor pairs ($i_1=i$
and $i_2=i+1$ in Eq.~\ref{Delta}).  Fig.~\ref{1ddata}(a) compares
$|P(r)|$ calculated by DMRG and CPMC.  The CPMC results were for
$\Delta\tau$=0.05 and a BP imaginary time of $\beta$=4. The CPMC
agrees very well with the exact DMRG result over several orders of
magnitude.  As with any Monte Carlo method, statistical error limits
the achievable accuracy of CPMC. In CPMC the most important parameter
governing the size of statistical fluctuations is the interaction
strength.  Fig.~\ref{1ddata}(b) shows a similar calculation to
Fig.~\ref{1ddata}(a), but with stronger interactions ($U$=8). Despite
a longer total projection length (5$\times$10$^5$), statistical
fluctuations are much more apparent in 
Fig.~\ref{1ddata}(b). However, within statistical error, 
results for larger $U$ still reproduce the exact correlations.

Back propagation adds an additional adjustable parameter $\beta$.  In
Fig.~\ref{1ddata}(c)-(d) we show $P(r)$ for a 32 site system with
$U$=4 as a function of $\Delta\tau$ and $\beta$. In our 1D tests, BP
converged smoothly with $\beta$ (note that the variation with $\beta$
in Fig.~\ref{1ddata}(c)-(d) would not be visible on the scales of
Fig.~\ref{1ddata}(a)-(b)). These results serve as a sensitive
verification of our implementation of the CPMC algorithm. Additional
 algorithmic tests are given in Appendix B.

\subsection{Two leg ladders}
\label{ladders}

 \begin{figure}[tb]
   \centerline{\resizebox{3.2in}{!}{\includegraphics{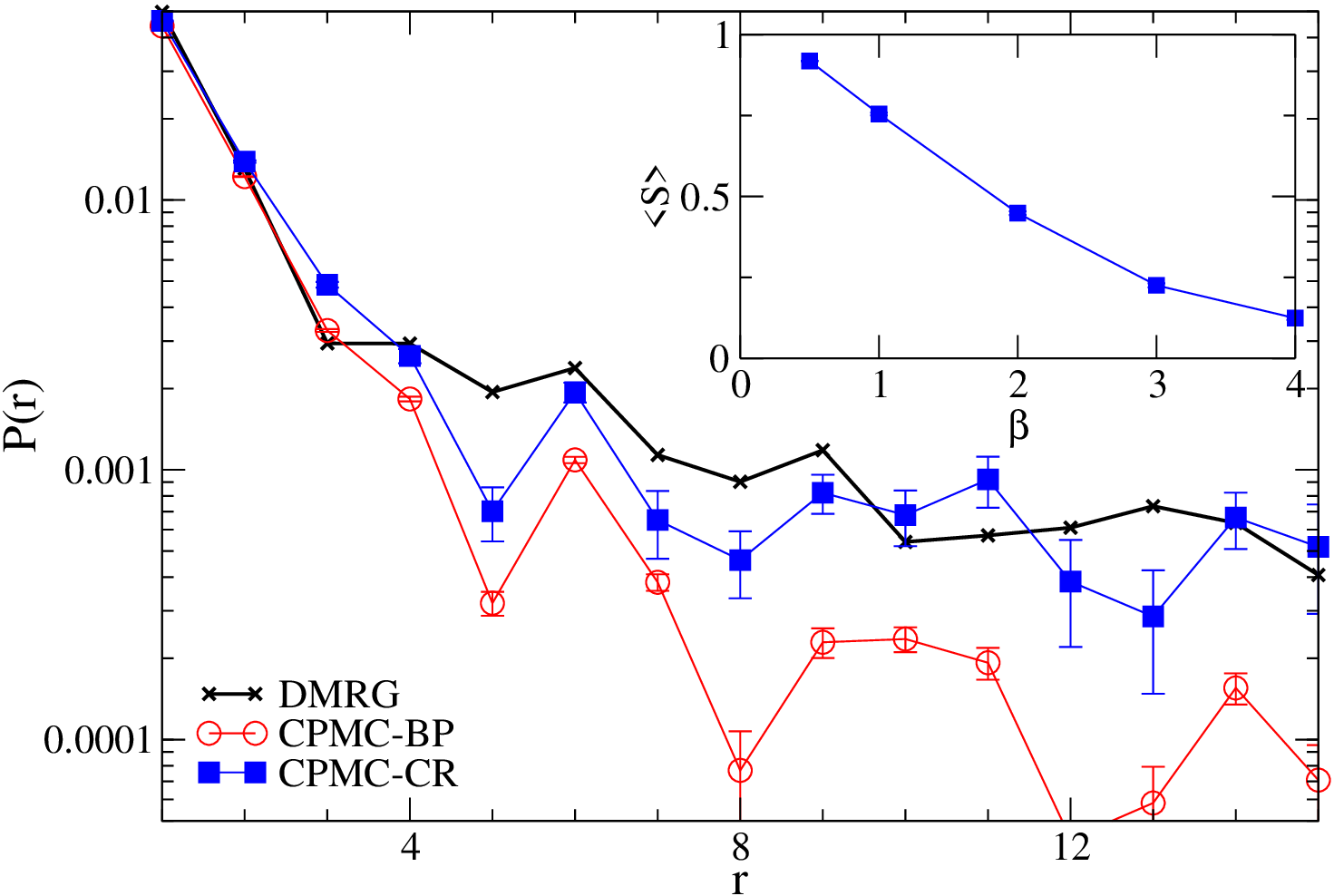}}}
   \caption{(color online) Pair-pair correlation $P(r)$ versus
     distance $r$ for a 32-leg ladder with open boundary conditions,
     56 particles, and $U$=4. The x's are numerically exact
     DMRG results, open circles are CPMC with BP,
      and filled squares are CPMC with CR. Both
     BP and CR used $\beta$=3. The inset
     shows  $\beta$ dependence of the average sign in the CR
      measurements. \label{ladderdata}}
 \end{figure}

 The two leg rectangular ladder is useful test case because while a
 sign problem does occur for QMC methods, highly accurate DMRG results
 are also available. Doped two-leg ladders are also known to have
 rung-singlet pair-pair correlations that decay with distance as power
 laws, with dominant pairing correlations for a lightly
   doped ladder \cite{Dolfi15a}. The
 appropriate pair creation operator is defined as Eq.~\ref{Delta},
 with sites $i_1$ and $i_2$ belonging to a single ladder
 rung. Fig.~\ref{ladderdata} shows $P(r)$ for an open boundary 32-rung
 ladder with 56 particles ($\frac{1}{8}$ doping) and $U=4$.  Here
 $P(r)$ is measured from the 8th rung of the lattice. We again use the
 noninteracting wavefunction for $|\Psi_t\rangle$.  For the CPMC
 calculations we used 6400 random walkers and $\Delta\tau=0.01$.  Each
 CR measurement used $\tau=\beta/4$ and 200 sweeps through the
 Hubbard-Stratonovich variables {\bf x}. The first 25 ``warm-up''
 sweeps were discarded before taking measurements
 (see Appendix \ref{appendixb}).  The average sign
 during the CR is shown in the inset of Fig.~\ref{ladderdata}.  As
 shown in Fig.~\ref{ladderdata}, CPMC using BP underestimates the long-range
 $P(r)$ in this system by nearly an order of magnitude, while CR
 produces much more accurate results. We did find the sign problem to
 have a significant effect however, and for stronger interactions
 ($U$=8) for this lattice, the average sign became too small to use
 CR.

\subsection{Square lattice}
\label{square}

The square periodic lattice has been studied by many authors using the
CPMC method \cite{Zhang97a,Zhang97b,Qin16a,Qin20a}.
Here we examine the  4$\times$4 square lattice which can be solved
exactly. We take $N_e$=10 particles, which is a closed-shell filling
in the noninteracting limit. For this system, the constrained path
approximation using the single-determinant free electron wavefunction for
$|\Psi_t\rangle$ is very accurate, with mixed energy
estimates accurate to a fraction of a percent (see Table \ref{4x4table}).  In this
calculation we used 1600 random walkers and $\Delta\tau$=0.01 to
minimize Trotter error (see Appendix \ref{appendixb}).
\begin{table}
   \begin{center}
     \begin{tabular}{|c|c|c|c|c|}
       \hline
       U & trial function &E$_{\rm var}$ \% err & $\bar{P}^{(t)}$ \% err  &E$_{\rm mix}$ \% err  \\
       \hline
       \hline
       4 & free electron & 9.4 & 27.3 & -0.0018(4)  \\
       \hline
       8 & free electron & 34.3 & 88.5 & -0.0415(7)  \\
       \hline
       8 & QP-PIRG & 3.1 & 22.5 & -0.0044(3)   \\
       \hline
     \end{tabular}
   \end{center}
   \caption{Relative errors in the variational energy,
     $E_{\rm var}=\langle \Psi_t|\hat{H}|\Psi_t\rangle$,
$\bar{P}^{(t)}=\langle \Psi_t|\bar{P}|\Psi_t\rangle$, and
     mixed energy
     estimates for free electron and QP-PIRG trial wavefunctions for
     the 4$\times$4 lattice with 10 particles. CPMC 
     results used $\Delta\tau$=0.01.  Numbers in parentheses are
     estimated uncertainties. \label{4x4table}}
\end{table}

Besides the free electron determinant, we tested a second trial
wavefunction constructed using the PIRG
 method \cite{Imada00a,Kashima01a,Mizusaki04a}. PIRG is a variational
method in which the ground state wavefunction is approximated by a 
finite sum of $L$ Slater determinants:
\begin{equation}
|\Psi_0\rangle \approx \sum_{i=1}^L c_i |\phi_i\rangle.
\end{equation}
The PIRG algorithm optimizes $c_i$ and the coefficients within each
determinant $|\phi_i\rangle$ through repeated applications of
$\exp(-\Delta\tau \hat{H})$, selecting the Hubbard-Stratonovich field
$x_i$ for each interaction that minimizes the variational energy
\cite{Imada00a,Kashima01a}.  The finite basis size can be corrected for by
repeating the calculation for increasing $L$ and extrapolating
expectation values versus the energy variance $\Delta E = (\langle
\hat{H}^2\rangle - \langle\hat{H}\rangle^2)/\langle\hat{H}\rangle^2$.
Advantages of PIRG are that it is exact for $U=0$, it has no sign
problem, and it is variational. The primary
disadvantages are the rapid increase of the $L$ needed to describe the
wavefunction with increasing
$U$, and poor computational scaling with respect to system size.
\begin{figure}[tb]
  \centerline{\resizebox{3.2in}{!}{\includegraphics{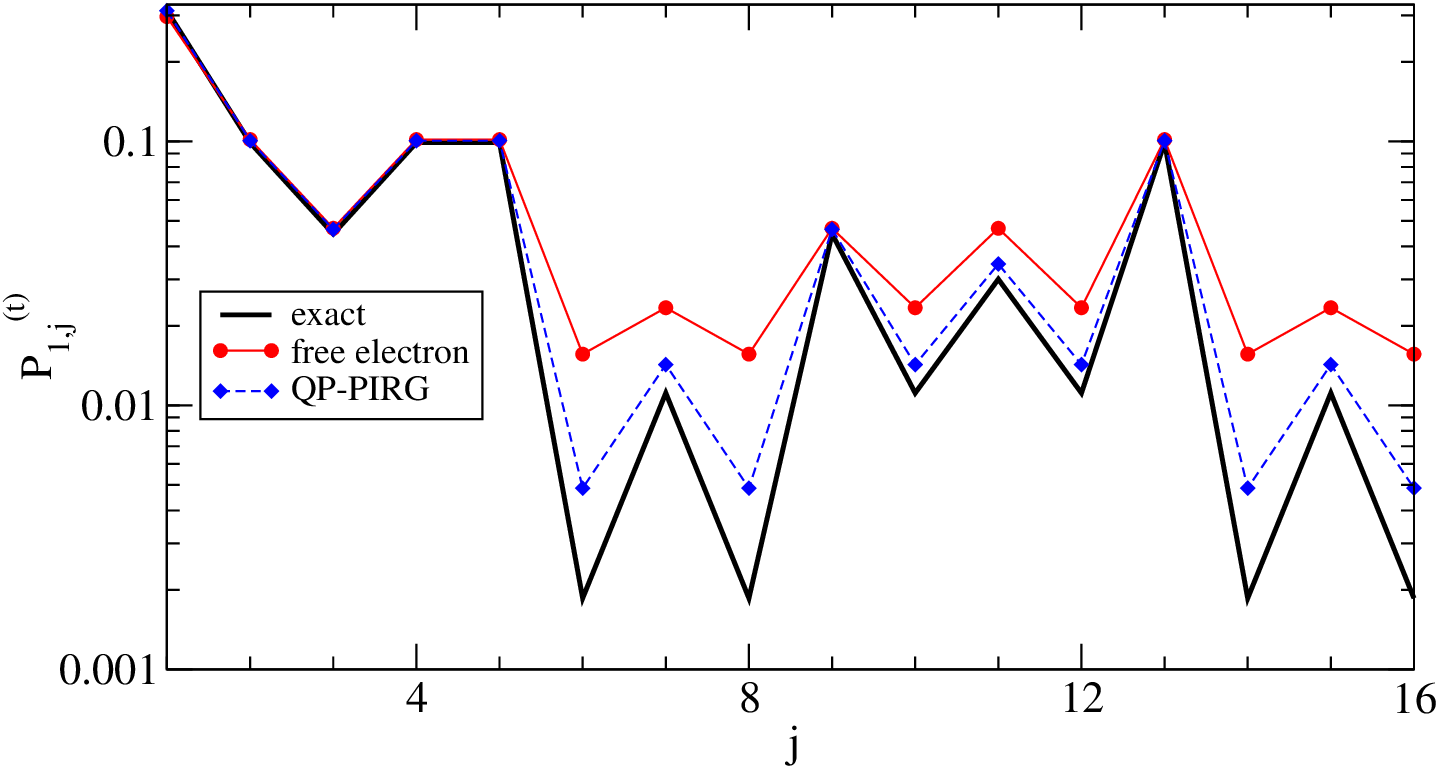}}}
  \caption{(color online) Pair-pair correlation between
      site 1 and each site $j$ of the lattice for for the 4$\times$4
      lattice with 10 electrons and $U$=8.  The thick line is the
      exact solution. Data points are the expectation value of $P(r)$
      in the trial wavefunction, with circles the free electron
      wavefunction and diamonds the $L=1$ QP-PIRG wavefunction.}
  \label{trial_pp}
 \end{figure}
\begin{figure}[tb]
  \centerline{\resizebox{3.2in}{!}{\includegraphics{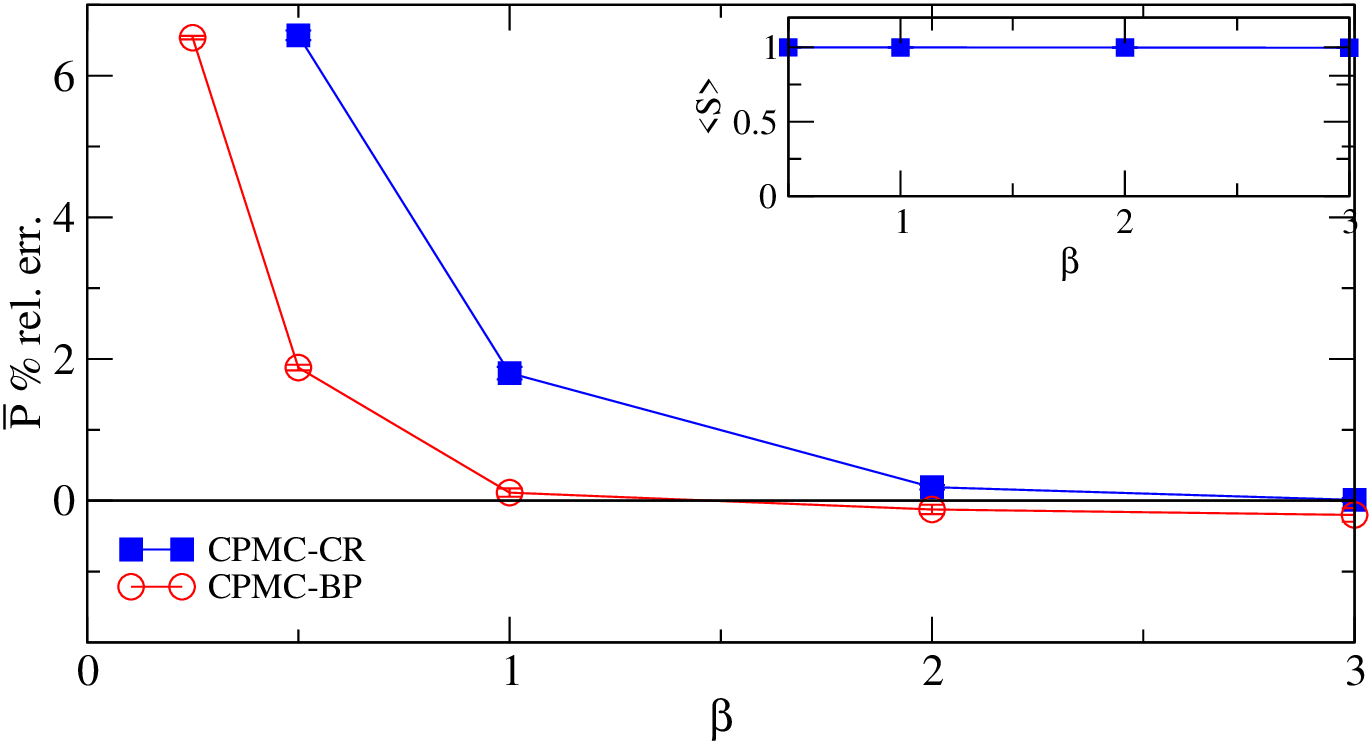}}}
  \caption{(color online) Percent relative error in the average
    long-range $d_{x^2-y^2}$ pair-pair correlation $\bar{P}$ versus
    $\beta$ for the 4$\times$4 lattice with 10 electrons, $U$=4, and
    $\Delta\tau$=0.01.  Circles and filled squares are BP and CR estimates
    with $\tau=\beta/2$, respectively. Both used the free electron
    trial function. The inset shows the average sign for the CR
    estimate. \label{4x4u4data}}
\end{figure}
\begin{figure}[tb]
  \centerline{\resizebox{3.2in}{!}{\includegraphics{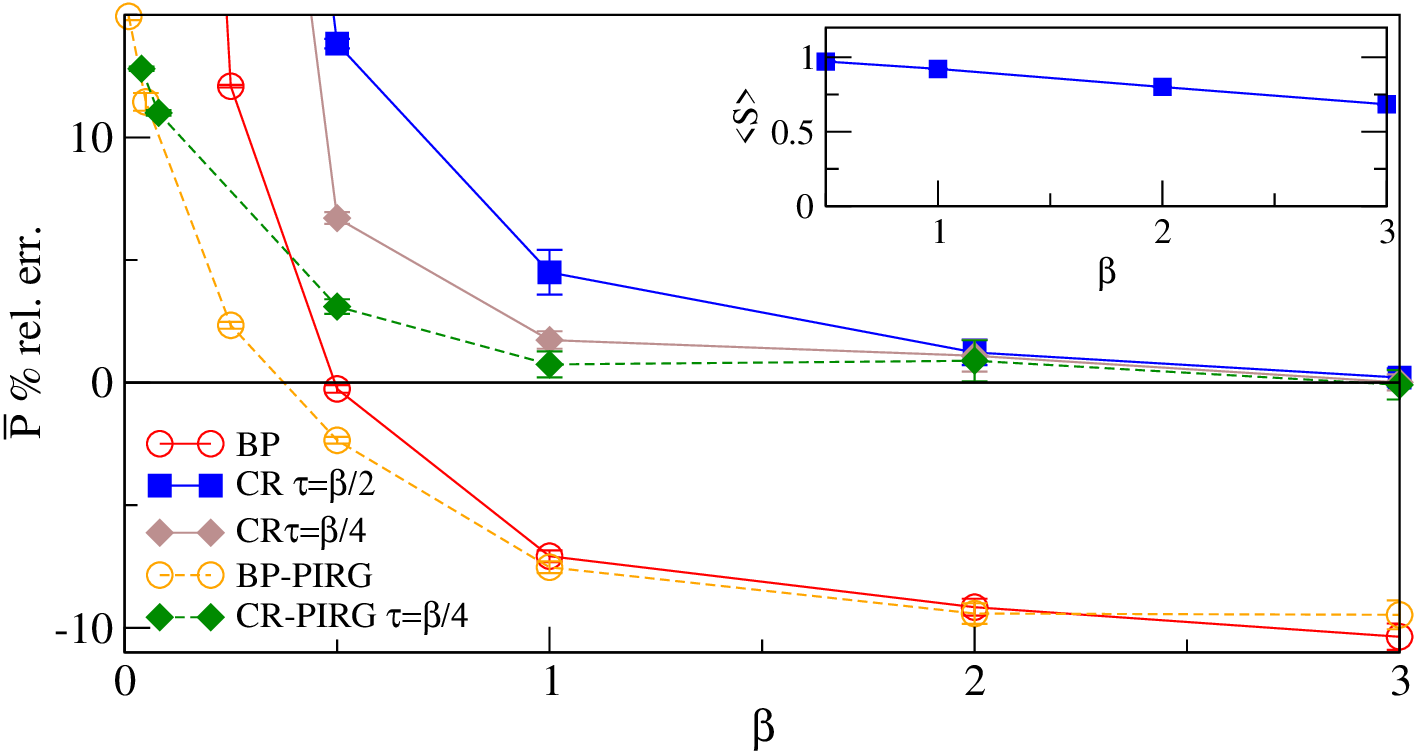}}}
  \caption{(color online) $\bar{P}$ versus $\beta$ for for the
    4$\times$4 lattice with 10 electrons, $U$=8, and
    $\Delta\tau$=0.01. Open (filled) symbols are BP (CR) estimates.
    Filled squares are CR  estimates with $\tau=\beta/2$, while
    filled diamonds used   $\tau=\beta/4$. Dashed lines indicate the  QP-PIRG
    rather than free
    electron trial function.  The inset shows the average sign for the
    CR estimate using the free electron trial function.\label{4x4u8data}}
 \end{figure}

A significant improvement is to
incorporate spin and spatial symmetries \cite{Mizusaki04a}. In this
``QP-PIRG'' approach, projection operators are used to symmetrize the calculation
of expectation values and overlaps between two determinants
\cite{Mizusaki04a}. For the square lattice we used translational and D$_4$
point group symmetries.   We also imposed spin parity symmetry, which
separates states into either even or odd total spin $S$.
The effective number of determinants in the PIRG
  wavefunction is $L$ multiplied by the number of symmetries.
  For the 4$\times$4 lattice with $L=1$, translational
  symmetry, D$_4$ point group symmetry, and spin parity, this
  is 256 determinants.
As shown in Table
\ref{4x4table}, this improves the variational energy of
$|\Psi_t\rangle$ and the mixed energy estimate by an order of
magnitude.
Long-range correlation functions are also more accurate in the QP-PIRG trial
function.
Fig.~\ref{trial_pp} plots
$\langle \Psi_t|P(r)|\Psi_t\rangle$ for
the $d_{x^2-y^2}$ pair-pair correlation
  between site 1 and each lattice site on the lattice for $U=8$.
 In Fig.~\ref{trial_pp} sites are numbered
  sequentially, with site 11 the furthest distance from site 1.
  Both trial functions are very accurate for short-range
  correlations, but the QP-PIRG wavefunction is significantly
  more accurate for long distance correlations. $\bar{P}$,
  the average of $P(r)$ for $r>2$, is
  also more accurate as well (see Table \ref{4x4table}).
  We  find that the average overlap of random walkers with the
  trial state, $\langle \Psi_t|\phi_i\rangle$, is larger for the QP-PIRG
  wavefunction, which is consistent with it being a better approximation
  to $|\Psi_0\rangle$ (see Appendix \ref{appendixb} Fig.~\ref{bp-overlap}).

 Fig.~\ref{4x4u4data} shows $\bar{P}$ as a function of
  $\beta$ for $U$=4 as estimated within CPMC using BP and CR.  These
results used the free electron wavefunction for $|\Psi_t\rangle$. For
$U$=4 and $\beta\agt$1 the BP estimate is quite accurate, but when
examined closely trends below the exact value as $\beta$ increases.
For $U$=8 the relative error in the mixed energy using the free
electron trial function is still much less than 1\% (see Table
\ref{4x4table}), but the error in the BP estimate of $\bar{P}$ is far
larger (see Fig.~\ref{4x4u8data}).  These results also show the
importance of considering the $\beta$ dependence of BP measurements.
BP underestimates pairing by around 10\% in this case for large
$\beta$.  For small $\beta$, BP overestimates pairing (BP and CR
become equivalent to the mixed estimator for $\beta\rightarrow 0$).
While BP produces the exact answer for $\beta\sim\frac{1}{2}$, it is
impossible in the general case to predict the precise $\beta$ giving
the most accurate results.  Compared to BP, we find that the CR
estimate of $\bar{P}$ converged to within statistical error of the
exact value for $\beta\sim 3$ for both $U$=4 and $U$=8.  As in Section
\ref{ladders}, each constrained release measurement used 200 total
sweeps.  In the CR measurement the average sign does fall with
increasing $\beta$, but in this case remained large enough that
essentially exact results can be obtained at $U$=8.

In Fig.~\ref{4x4u8data} we show two different choices for $\tau$ (see
Eq.~\ref{cr}), either $\tau=\beta/2$ or $\tau=\beta/4$. We found that $\tau\approx\beta/4$
produced more accurate results which converged
slightly faster with increasing $\beta$.  This
can be understood from Eq.~\ref{zerot}, where in the constrained
release measurement $|\Psi_L\rangle=|\Psi_t\rangle$ and
$|\Psi_R\rangle=|\phi^0_i\rangle$. Because $|\phi^0_i\rangle$ is
expected to be a better representation of $|\Psi_0\rangle$ than
$|\Psi_t\rangle$, it is then better to perform a longer imaginary time
projection in Eq.~\ref{zerot} to the left of $\hat{\mathcal O}$ than
to its right.  Optimizing $\tau$ can be a significant improvement when
using CR, especially in cases where the maximum $\beta$ attainable
is limited by the sign problem.

An important consideration is whether a ``better''
  $|\Psi_t\rangle$ improves correlation functions calculated within
  CPMC.
In Fig.~\ref{4x4u8data} we also plot BP and CR
estimates of $\bar{P}$ using the QP-PIRG $L=1$ wavefunction as
$|\Psi_t\rangle$.
BP and CR estimates both improve at small $\beta$ with the QP-PIRG trial
function. This is
expected, since in the mixed estimate limit of $\beta\rightarrow0$
both $\langle\Psi_t|$ and $|\phi_i\rangle$ should be more accurate
representations of $|\Psi_0\rangle$.  Surprisingly, for larger
$\beta$, the improved trial function did not improve the BP
results.
We will revisit this issue in Section \ref{5site}.
The more accurate trial function did slightly improve the CR
results, but at the expense of a significantly longer calculation.

\subsection{Anisotropic triangular lattice}
\label{aniso}

\begin{table}
  \begin{center}
    \begin{tabular}{|c|c|c|c|}
      \hline
      U & QP-PIRG & $L$=1 $E_{\rm var}$ \% err. & CPMC \\
      \hline
      \hline
      1 & -53.44055(2) & 0.05722(4) & -53.44054(1) \\
      \hline
      2 & -45.705(2) & 0.708(4) & -45.70346(4) \\
      \hline
      3 & -38.810(5)  & 2.93(1) &-38.7891(1) \\
      \hline
      4 & -32.80(2)  & 8.06(6) &-32.7108(3) \\
      \hline
      5 & -27.502(8) & 13.70(3) & -27.511(1) \\
      \hline
    \end{tabular}
  \end{center}
  \caption{Ground state energy  versus $U$ for the 6$\times$6
 half-filled   anisotropic triangular lattice with $t^\prime$=0.5.
 The second column are  variance-extrapolated  QP-PIRG results
using $L$ up to 1024,
column three  the QP-PIRG  error for $L=1$ 
  relative to the extrapolated QP-PIRG value, and column four
  CPMC results using the $L$=1 QP-PIRG wavefunction for $|\Psi_t\rangle$.
CPMC results  used $\Delta\tau$=0.01 and the mixed estimator. QP-PIRG results are
    variance extrapolated with a linear least-squares fit using the three
    largest $L$ wavefunctions. Numbers in
    parentheses are estimated uncertainties.\label{6x6table}}
\end{table}
  
  \begin{figure}[tb]
  \centerline{\resizebox{3.2in}{!}{\includegraphics{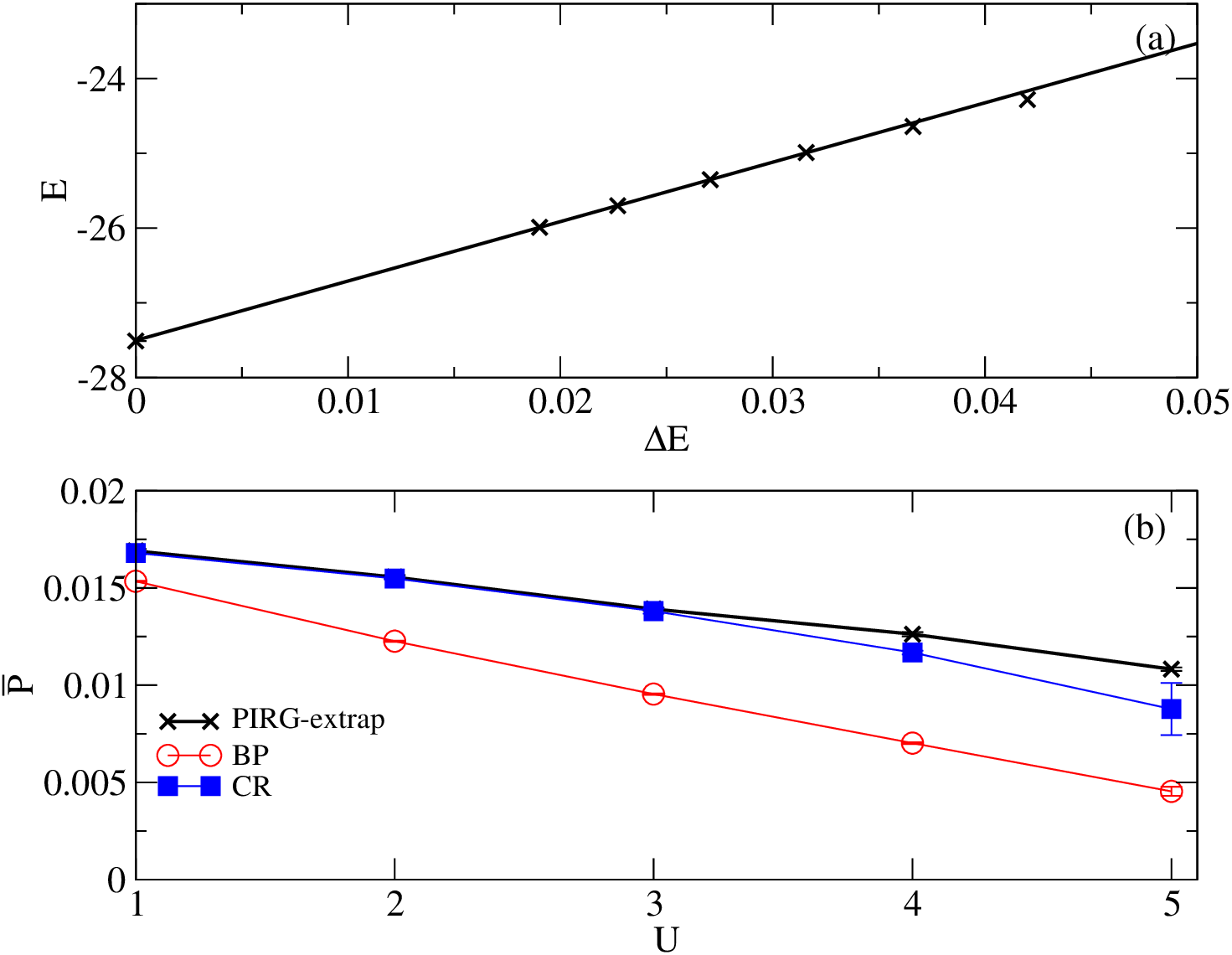}}}
  \caption{(color online) (a) Variance extrapolation of the energy
    using the QP-PIRG method for the half-filled 6$\times$6 anisotropic lattice
    with $t^\prime=0.5$ and $U$=5.  The line is a linear
    least-squares fit.  (b) Average long-range $d_{x^2-y^2}$ pair-pair
    correlation versus $U$ on the 6$\times$6 anisotropic triangular
    lattice with $t^\prime$=0.5. X's, circles, and filled squares
    were calculated using extrapolated QP-PIRG, CPMC with BP, and CPMC with CR,
    respectively. CPMC results used $\Delta\tau=$0.05.}
  \label{6x6}
\end{figure}  

The anisotropic triangular lattice consists of a square lattice with
nearest-neighbor bonds $t$, with an additional next-nearest-neighbor
bond $t^\prime$ across one of the diagonals of each plaquette of the
lattice. The limit $t^\prime$=$t$ corresponds to the triangular
lattice. The half-filled Hubbard model on this lattice has been used
to model the conducting layers of some organic charge-transfer solid
superconductors (see for example \cite{Kanoda11a}).  At half filling
the non-interacting band is degenerate for square periodic clusters,
so a multi-determinant $|\Psi_t\rangle$ is required to preserve the
translational invariance of CPMC results.  We again used a QP-PIRG
trial function using translations, $C_{2v}$ point group symmetries,
and spin parity ($8N$ total symmetries).  Extrapolated QP-PIRG results
allow comparison  to essentially exact results for systems larger
than can be solved exactly \cite{Dayal12a}.  For a 6$\times$6 lattice
with $t^\prime$=0.5 and  $U \leq 5$ we used QP-PIRG with $L$ up to
1024.  A typical extrapolation of the ground state energy versus
energy variance is shown in Fig.~\ref{6x6}(a), where data points
correspond to $L$ of 1024, 512, 256, 128, 64, and 32.  For CPMC, we
again used the $L$=1 QP-PIRG wavefunction for $|\Psi_t\rangle$. For
the 6$\times$6 lattice this trial function is composed of 288 total
determinants. The table in Fig.~\ref{6x6} compares energy
  results 
  for extrapolated QP-PIRG, QP-PIRG with $L=1$, and CPMC using the
$L=1$ QP-PIRG wavefunction as $|\Psi_t\rangle$.

Fig.~\ref{6x6}(b) shows results for the $d_{x^2-y^2}$ $\bar{P}$ as a
function of $U$. $\bar{P}$ decreases continuously with increasing $U$,
implying that the ground state does not have superconducting order
\cite{Dayal12a}. As in Section \ref{square}, we found that the CPMC
BP estimate had converged for $\beta\agt$2; in Fig.~\ref{6x6}(b) we used
$\Delta\tau$=0.05 and $\beta=2$.  The CR results used $\beta=2$ and
$\tau=\beta/4$ and agreed very well with extrapolated QP-PIRG.
Compared to QP-PIRG, BP again underestimated $\bar{P}$, with the
discrepancy growing with increasing $U$.

\subsection{Exact versus approximate trial functions}
\label{5site}

  \begin{figure}[tb]
  \centerline{\resizebox{3.2in}{!}{\includegraphics{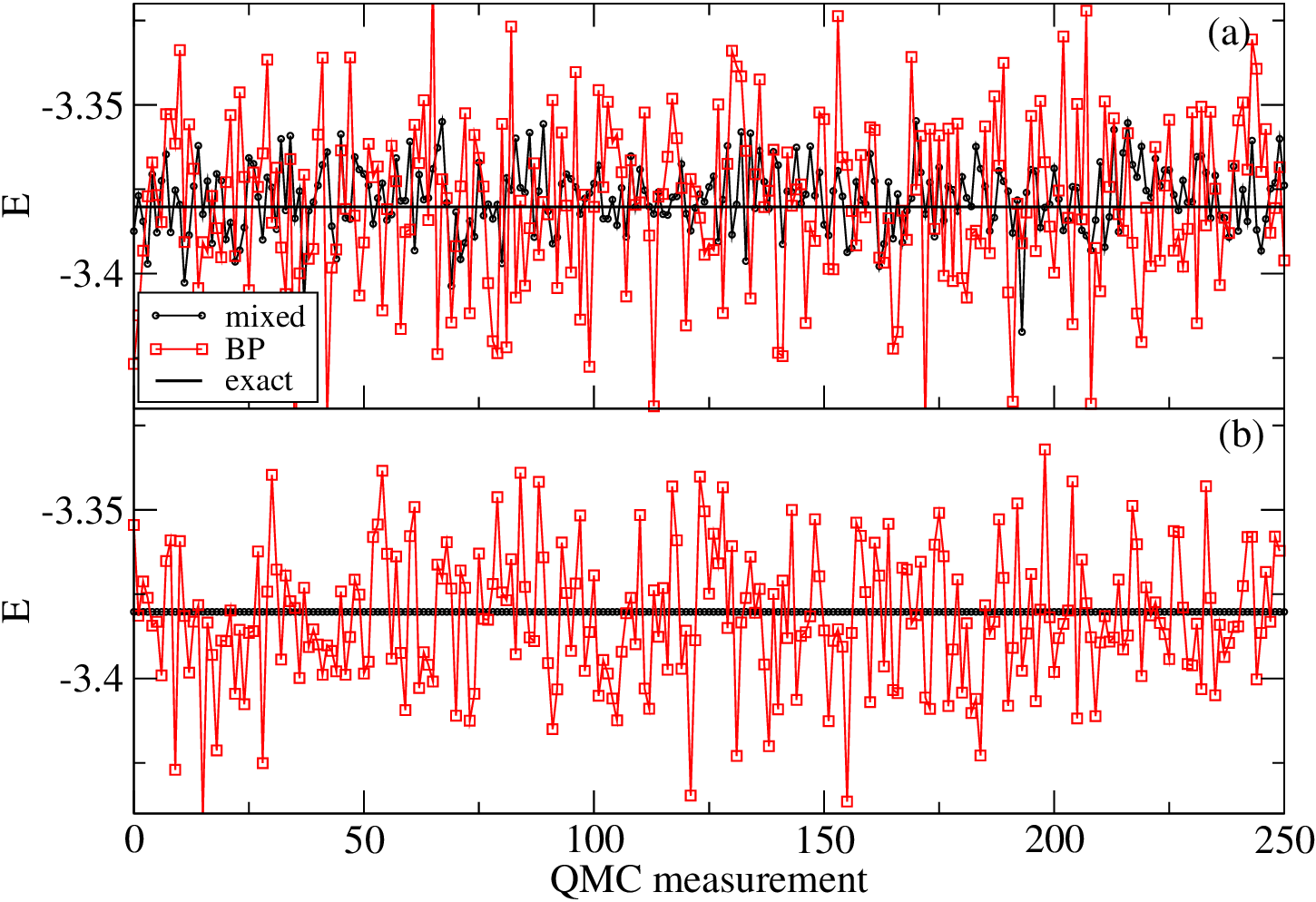}}}
  \caption{(color online) (a) Mixed and BP energy measurements for a five site
    ring with anti-periodic boundary conditions, $N_p$=4, $U$=8, $\Delta\tau$=0.01,
    and $\beta=2$. The noninteracting wavefunction
    is used for $|\Psi_t\rangle$.  (b) Same as (a), except $|\Psi_t\rangle$ is
    the exact ground state wavefunction.}
  \label{5s-apbc}
\end{figure}  
  \begin{figure}[tb]
    \centerline{\resizebox{3.2in}{!}{\includegraphics{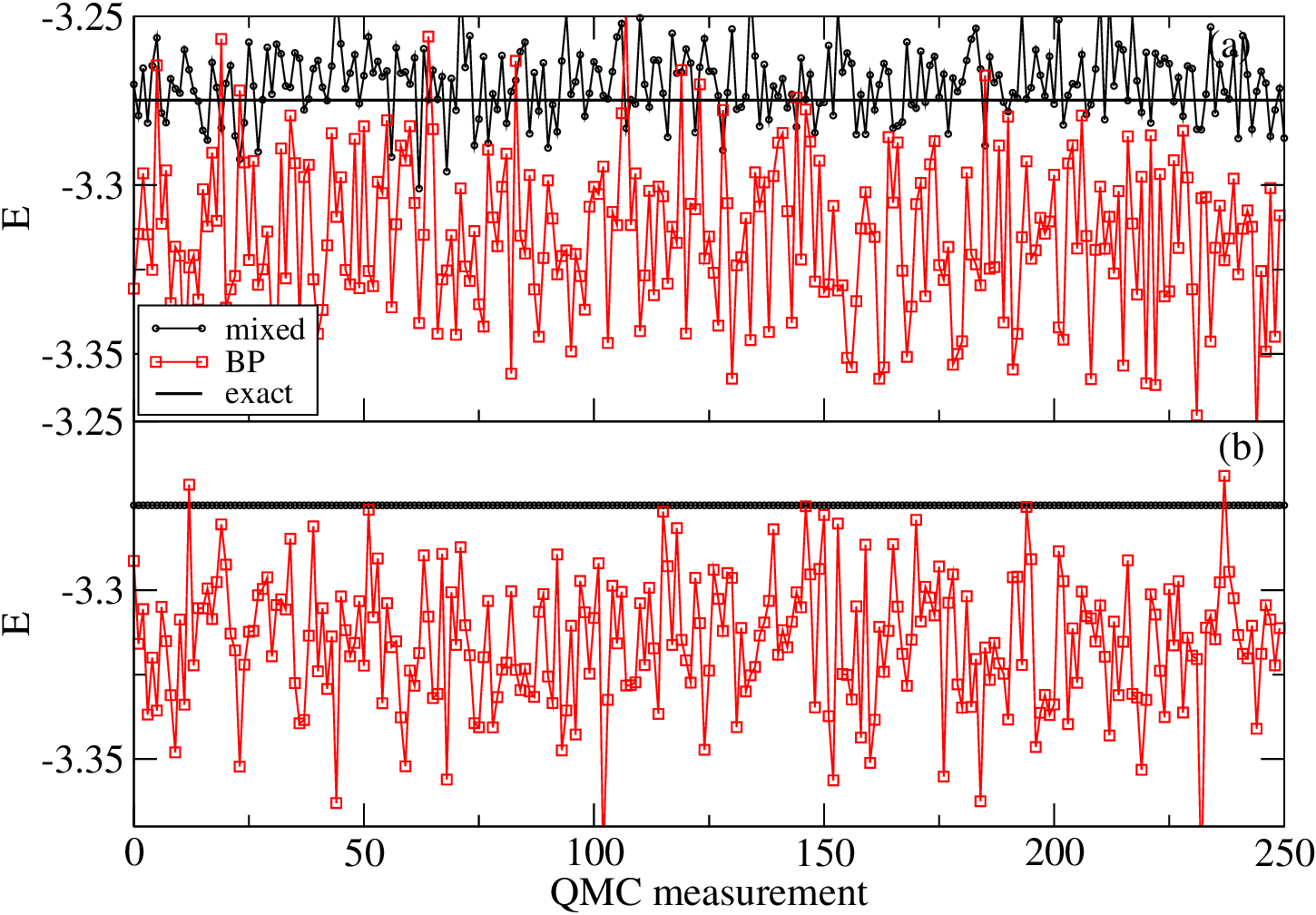}}}
  \caption{(color online) (a) Mixed and BP energy measurements for a five site
    ring with periodic boundary conditions and one long-distance hopping (see text), $N_p$=4, $U$=8, $\Delta\tau$=0.01, and $\beta$=2. The noninteracting wavefunction
    is used for $|\Psi_t\rangle$.  (b) Same as (a), except $|\Psi_t\rangle$ is
the exact ground state wavefunction.}
  \label{5s}
\end{figure}  

To clarify the errors in BP measurements, we present one additional
set of calculations. Fig.~\ref{5s-apbc} shows energy measurements for
a five site ring with four particles (two spin up and two spin down),
anti-periodic boundary conditions, and $U$=8. The $x$ axis in the
figure corresponds to the sequence of Monte Carlo measurements, with
10 binned measurements for each data point.  Because a single BP
measurement involves many imaginary time steps, the total imaginary
time projection for the mixed versus BP plots are not equivalent. The
goal in these figures is to show the amount of fluctuation in the
measurements.  The results in Fig.~\ref{5s-apbc}(a) used the
free-electron wavefunction as $|\Psi_t\rangle$.  The averages of the
measurements shown in Fig.~\ref{5s-apbc}(a) are only slightly
different from the exact value due to the finite $\beta$ and
$\Delta\tau$.  As in Section \ref{1d}, we find that both the mixed and
BP energy estimates in this 1D system are consistent with the exact
energy, assuming proper extrapolation in $\Delta\tau$ and $\beta$.

Using PIRG, for a small system it is feasible to construct the {\it
  exact} ground-state wavefunction out of Slater determinants. Using
12 determinants, the PIRG energy matched the exact energy to eight
decimal places; correlation functions also matched exact results.
 When the exact wavefunction is used in CPMC as $|\Psi_t\rangle$, because of
importance sampling the mixed estimator is equal to the exact energy with
no fluctuations \cite{Zhang97a}.  In our tests,  mixed energy
estimates remained constant to within 10$^{-7}$ as shown in
Fig.~\ref{5s-apbc}(b) (in Fig.~\ref{5s-apbc}(b) the ``exact'' line is
hidden under the ``mixed'' points). The average of BP measurements of
the energy in Fig.~\ref{5s-apbc}(b) is still consistent with the exact
energy, but importantly, the scatter of the BP measurements is
essentially the same
 (the average and estimated uncertainty of 250 BP bins
 was -3.3818$\pm$0.0017 in Fig.~\ref{5s-apbc}(a)
versus -3.3820$\pm$0.0013 in Fig.~\ref{5s-apbc}(b)).

Fig.~\ref{5s} shows results for a similar system where a sign problem
is present. Here we take a five site periodic ring with one added
long-range hopping of magnitude 0.7$t$ between sites two and four. The
exact ground state for this system is a spin singlet; the
noninteracting ground state is non-degenerate.  In Fig.~\ref{5s}(a) we
use the noninteracting ground state for $|\Psi_t\rangle$.  As shown in
Fig.~\ref{5s}(a), the averages of both mixed and BP energy estimates are further
from the exact energy, and do not converge to the exact value
if an extrapolation in $\beta$ and $\Delta\tau$ is performed (not shown here).
 For this system PIRG can also calculate
the exact ground state. Results using the exact wavefunction for
$|\Psi_t\rangle$ are shown in Fig.~\ref{5s}(b). Here again, the mixed
energy estimate is constant and equal to the exact energy. The average
of BP estimates is however {\it not} equal to the exact energy.  The
average of the BP energies was -3.3152$\pm$0.0015 in Fig.~\ref{5s}(a)
and -3.3155$\pm$0.0012 in Fig.~\ref{5s}(b), again with only possibly a slight
decrease in uncertainty when $|\Psi_t\rangle=|\Psi_0\rangle$.

As shown in Fig.~\ref{5s-apbc} and \ref{5s}, fluctuations in BP
measurements are far greater than in mixed measurements, even if
$|\Psi_t\rangle$ is the exact wavefunction.  One exception is for
$U$=0, where BP trivially gives exact results. Fluctuations in BP
measurements rapidly grow with increasing $U$.  Recall that the
purpose of BP is to approximate $\langle \Psi_0|$ in the expectation
value $\langle \Psi_0 | \hat{\mathcal{O}} | \Psi_0 \rangle$ as a
distinct state from $|\Psi_0\rangle$. $|\Psi_0\rangle$ is estimated by
the CPMC importance-sampled random walk (see Section
\ref{mixedsection}).  $\langle \Psi_0|$ is built up from the trial
wavefunction as $\langle \Psi_t|e^{-\beta\hat{H}}$ (Eq.~\ref{bp}). The
Hubbard-Stratonovich fields in $e^{-\beta\hat{H}}$ however are taken
from the forward-going importance sampled random walk. When these
fields are applied to an arbitrarily chosen starting state
$\langle\Psi_t|$, the resulting imaginary time propagation does not
benefit from importance sampling, which depends on {\it both initial
  and final Slater determinants} (Eq.~\ref{importance}).  Arbitrarily
using the importance-sampled fields on different Slater determinants
effectively biases $P(x_i)$ in an uncontrolled manner.  While
Reference \cite{Zhang97a} claimed that BP is exact when
$|\Psi_t\rangle=|\Psi_0\rangle$, arguments there did not account for
this issue.  The result is large fluctuations in BP measurements;
while an improved $|\Psi_t\rangle$ reduces fluctuations in the
estimate of $|\Psi_0\rangle$, it does not significantly reduce
fluctuations in the BP estimate of $\langle \Psi_0|$. This is the
reason that improved trial functions do not improve BP accuracy, and
why BP estimates are approximate even when the exact wavefunction is
used as $|\Psi_t\rangle$.

\section{Conclusions}
\label{conclusions}

The calculation of correlation functions using CPMC depends on
additional approximations beyond the assumption of a trial
wavefunction to control the sign problem. The main focus of our
work is to critically examine the effect of these approximations.
Below we summarize the  main conclusions of our work.

First, we have clarified the conditions under which CPMC is
statistically exact for one dimensional systems. CPMC is exact for
uniform one-dimensional systems with periodic boundaries and odd
$N_\sigma$ or anti-periodic boundaries and even $N_\sigma$.  In these
cases, both the mixed estimate of the energy and BP estimates of
correlation functions converge to the exact values as a function of
$\Delta\tau$ and $\beta$.  The exactness in 1D can be understood based
on the sign properties of one dimensional wavefunctions and the
sign-preserving nature of the imaginary-time propagator.  Unlike
several other QMC methods, CPMC is in general not exact for 1D systems with
open boundaries. This appears to be the case whenever the electron
density is non-uniform. Further work to more precisely determine the
required conditions is of interest, as it may help to understand
 the behavior of the sign problem in CPMC in more complicated
cases.

We have investigated the properties of the BP estimator
in a variety of cases.  We have shown that BP tends to underestimate
long-range superconducting pair-pair correlations, and in general
becomes less accurate as $U$ or the range of the correlation
increases.  In general, BP is only exact in the limit of $U$=0 or in
special cases like 1D.  BP becomes exact in 1D
because there the imaginary time propagator does not generate minus
signs.  While we have focused on pair-pair correlations, because all
measurements in CPMC are based on estimates of the Green's function
\cite{Gubernatis16a}, similar errors will be seen in other correlation
functions.  Further important conclusions of our work are that
improving the quality of the trial function does not necessarily
improve the BP estimate, and more importantly, BP results are not
exact even when the exact wavefunction is used for
$|\Psi_t\rangle$. While improved trial functions decrease the
fluctuations in the CPMC importance-sampled random walk making the
mixed energy estimate more accurate, they do not reduce fluctuations
in BP measurements when the same propagators are applied to $\langle
\Psi_t|$. We suggest that these uncontrolled fluctuations result in
partial cancellation in long-range measurements, which is the source
of the underestimate of long-range pairing.  It would be interesting
to investigate how BP errors appear in CPMC in models with complex
propagators, where a more complicated phase problem is introduced by
the overlap constraint.  For these cases variants of the BP algorithm,
for example path restoration, have been proposed \cite{Motta17a}.

The CR technique can be statistically exact, provided the sign problem
does not prevent its use.  In our tests for the two-leg ladder, CR
calculations were not possible with $U\agt 6$ due to the sign problem.
On the other hand, the region of interesting physics in the
anisotropic triangular lattice for $t^\prime$=0.5 is near the
metal-insulator transition at $U\sim 5$ \cite{Dayal12a}, which as we
show in Section \ref{aniso}, can be accessed using CR. Furthermore,
while it is computationally expensive, our results show that improving
the trial function does lead to more accurate results with CR.
Because CR is far slower than BP, further work in optimizing the
additional parameters it introduces (for example the number of sweeps
over ${\bf x}$, the number of warm-up sweeps, and $\tau$) would be
useful. An additional technique we did not explore was the possibility
of only releasing part of the fields ${\bf x}$, keeping the rest at
their initial value from the CPMC path \cite{Xiao23a}. Such a
measurement would be essentially a hybrid between BP and CR.
One disadvantage is that the portion of the path with fixed
fields would again lead to an uncontrolled error in measurements.

Regarding the search for SC in Hubbard-type models,
an alternate technique to direct calculation of superconducting
pair-pair correlations is to calculate the energy as a
function of an assumed pairing field \cite{Qin20a,Xu24a}.  While these
results have suggested long-range SC in the frustrated 2D lattice
\cite{Xu24a}, it would be desirable to confirm SC through calculation
of pair-pair correlations as well.  Our finding that BP tends to
underestimate pair-pair correlations suggests that prior CPMC
calculations of pairing should be revisited.

\acknowledgments

This work was supported  by the US National Science Foundation Division
of Materials Research under grant number 2348712.

\appendix
\section{CPMC in one dimension}
\label{appendixa}

In this Appendix we present more details on the observation that CPMC
is exact in 1D.
For simplicity we consider a 1D model of spinless fermions,
\begin{equation}
  \hat{H}=-\sum_{i=1}^Nt_{i,i+1}(c^\dagger_{i+1}c_i + H.c.) + V\sum_i n_{i+1}n_i.
  \label{hspinless}
\end{equation}  
In Eq.~\ref{hspinless}, $c^\dagger_i$ creates a fermion on site $i$,
and $n_i=c^\dagger_ic_i$. We assume the hopping integrals are uniform,
$t_{i,i+1}=t$, where $t$ is a positive quantity.  Choices
  for the boundary hopping ($t_{N,N+1}\rightarrow t_{N,1}$) in
Eq.~\ref{hspinless} include periodic($t_{N,1} = t$),
antiperiodic ($t_{N,1} = -t$), or open ($t_{N,1} = 0$).  CPMC appears
to be exact when $N_\sigma$ is odd (even) for periodic (antiperiodic)
boundary conditions, but {\it not} exact in 1D with open
  boundary conditions.  In all of these cases, the
non-interacting ground state is nondegenerate.  Similar restrictions
on boundary conditions and numbers of particles apply to other QMC
methods in 1D such as the world line method \cite{Hirsch82a}, Greens
Function Monte Carlo \cite{Lee84a}, the Stochastic Series Expansion
method \cite{Sandvik92a}, and Full Configuration Interaction Quantum
Monte Carlo \cite{Ghanem21a}. Below we will specifically address the
case of periodic boundary conditions.

We write the exact ground state wavefunction as
\begin{equation}
  |\Psi_0\rangle=\sum_i\chi_i|\chi_i\rangle.
  \label{psi0}
\end{equation}   
In Eq.~\ref{psi0}, $|\chi_i\rangle$ denotes a many-body electron
configuration and $\chi_i$ a real coefficient.  A key observation
 is that for a 1D periodic fermion system with odd
  $N_\sigma$ the coefficients $\chi_i$ can be chosen to be {\it all
  positive} (see for example Reference \cite{Lieb03a}).  Because this
is true for any strength of interaction, this further implies that the
overlap of $|\Psi_0\rangle$ and the noninteracting Slater determinant
ground state, $\langle\Psi_0|\phi_0\rangle$ can always be chosen to be
positive.

Calculations with CPMC have noted that for 1D systems, with the choice
of $|\Psi_t\rangle=|\phi_0\rangle$, the overlap
$\langle\Psi_t|\phi\rangle$ never becomes negative and walkers are
never removed by the constraint. Reference \cite{Carlson99a} showed
analytically that $\langle\Psi_0|\phi\rangle$ remains positive
following projection in imaginary time for a small system of two
spinless fermions on a 3-site lattice with open boundaries, with the
result that the CPMC random walk gave statistically exact results for
the ground state energy. Here we show that the overlap with the
noninteracting trial function remains positive under imaginary time
projection for a general 1D system.  A Hubbard-Stratonovich
transformation similar to Eq.~\ref{hs} can be applied to the $V$
interaction.

A random walker $|\phi\rangle$ starts with
$|\phi\rangle=|\phi_0\rangle$, which as noted above has a positive
overlap with the true ground state $|\Psi_0\rangle$. A consequence of
this is that the coefficients $\alpha^0_i=\langle\chi_i|\phi_0\rangle$
are also all positive.  In CPMC the walker is
modified by $\exp(-\Delta\tau \hat{H})$, which we write
here to first order in $\Delta\tau$,
\begin{equation}
|\phi^\prime\rangle =e^{-\Delta\tau \hat{V}}e^{-\Delta\tau \hat{K}}|\phi\rangle.
\end{equation}
The exponential of the hopping term can be further broken down using
the ``checkerboard breakup'' \cite{Gubernatis16a}, in which the matrix
$\exp(-\Delta\tau\hat{K})$ is approximated as a product of individual
matrices for each hopping between sites $i$ and $i+1$,
$$
e^{-\Delta\tau\hat{K}}\approx \prod_ie^{-\Delta\tau K_{i,i+1}}.
$$
This approximation becomes exact in the limit $\Delta\tau\rightarrow0$.
The matrix $\exp(-\Delta\tau K_{i,i+1})$ has $\cosh(\Delta\tau|t|)$
for its $(i,i)$ and $(i+1,i+1)$ diagonal elements
and $\sinh(\Delta|t|)$ for its
$(i,i+1)$ and $(i+1,i)$ off-diagonal elements, all of which are
positive quantities. Its remaining diagonal elements are all 1.
The Slater determinant $|\phi\rangle$ acted
 on by $\exp(-\Delta\tau\hat{K})$ will now have non-zero overlaps $\alpha_j$
 with configurations $|\chi_j\rangle$ related to $|\chi_i\rangle$ by
 hopping of electrons. However, the new scalar
 products
 $\langle\chi_i|\exp(-\Delta\tau\hat{K})|\phi\rangle$ will all
 remain positive because of the positivity of the matrix elements
 of $\exp(-\Delta\tau K_{i,i+1})$.

 The interaction term
 $\exp(-\Delta\tau\hat{V})$  following the
 Hubbard-Stratonovich transformation is represented by a diagonal matrix,
 whose diagonal elements are again all positive quantities (Eq.~\ref{hs}).
 The result is that following propagation of the Slater determinant
 by $\exp(-\Delta\tau\hat{H})$, the new coefficients $\alpha_i=\langle\chi_i|\phi^\prime\rangle$
 remain all positive, as does the overlap with the true ground
 state $\langle \Psi_0|\phi^\prime\rangle$. Thus the CPMC random walk  samples only
 determinants $|\phi\rangle$ for which $\langle\Psi_0|\phi\rangle>0$ and does
 not suffer from a sign problem.
 Because no walkers are ever
 removed from the ensemble due to the constraint, 
   and the reversed propagator also does not generate constraint
   violations,
BP  will also return statistically exact results, provided conditions
 such as convergence in $\beta$ are met.

  While many other QMC methods are exact when used for 1D
   systems with open boundary conditions, CPMC is not. While we are
   not able to rule out the possibility of the method being exact for
   certain specific open lattices and parameters, in tests we found
   small errors in the extrapolated mixed energy as well as BP
   calculations of correlations. In particular, we found errors in the
   charge density, which varies with position in a system with open
   boundaries. The variation of charge density with position is
   different for $|\phi_0\rangle$ compared to the true ground state
   $|\Psi_0\rangle$, which we speculate causes the propagation in
   imaginary time to no longer be restricted to regions where $\langle
   \Psi_0|\phi\rangle > 0$. 
 
\section{Further method tests}
\label{appendixb}

\begin{figure}[tb]
  \centerline{\resizebox{3.2in}{!}{\includegraphics{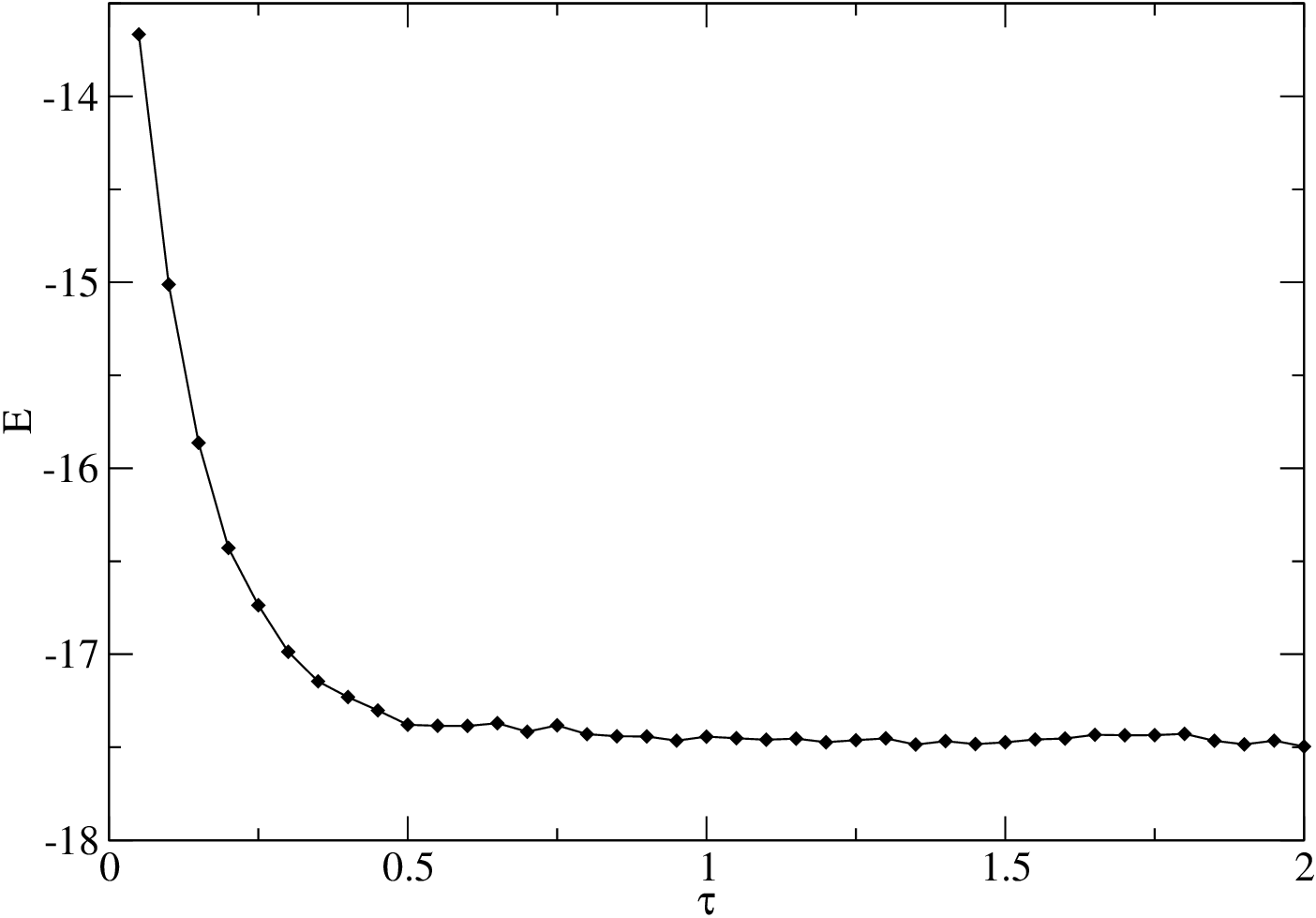}}}
  \caption{Mixed energy estimate versus length
    of projection in imaginary time. Data are for the 4$\times$4 lattice with
    $N_e$=10, $U$=8, $\Delta\tau$=0.05, and 1600 walkers.}
  \label{equilibration-convergence}
\end{figure}
\begin{figure}[tb]
  \centerline{\resizebox{3.2in}{!}{\includegraphics{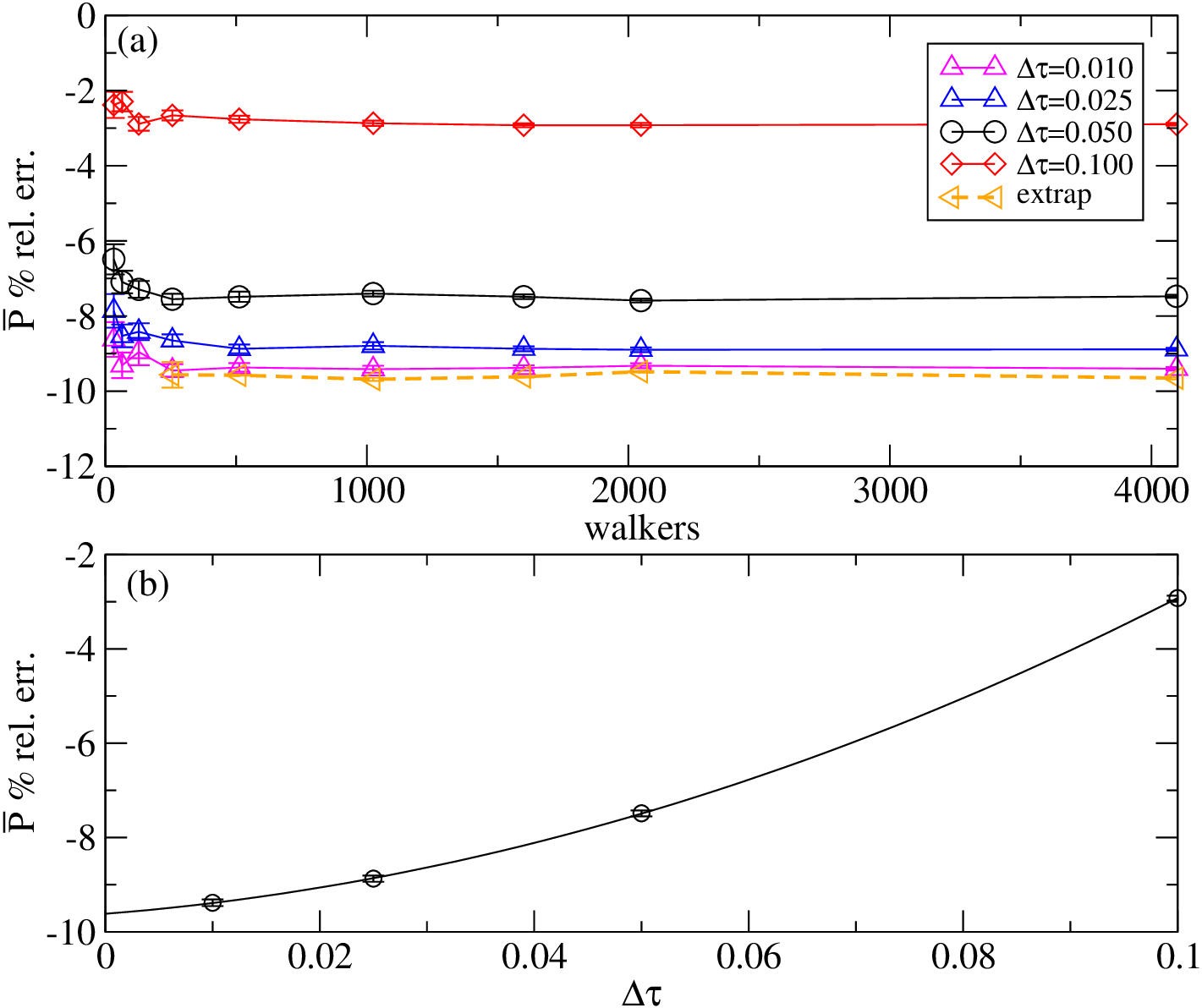}}}
  \caption{(color online) (a) Relative error in the $d_{x^2-y^2}$
    $\bar{P}$ compared to the exact value calculated using back
    propagation versus the number of random walkers and
    $\Delta\tau$. Data are for the 4$\times$4 lattice with $N_e$=10,
    $U$=8, and $\beta=2$. The dashed line shows the results
    extrapolated to $\Delta\tau$=0 using a least-squares quadratic
    fit. (b) The $\Delta\tau$ extrapolation for 1600 walkers.}
  \label{walker-convergence}
\end{figure}

In this Appendix we present additional tests on the convergence of the
method against various parameters.

In CPMC each random walker is initialized to $|\Psi_i\rangle$, and a
number of equilibration steps must be performed before taking
measurements.  Fig.~\ref{equilibration-convergence} shows the mixed
energy as a function imaginary time. Our results here are
similar to Reference \cite{Zhang97a}. While the mixed energy
converges quite rapidly, in the tests presented in this paper we
 take a much more conservative equilibration cutoff of $\tau\sim$100.
\begin{figure}[tb]
  \centerline{\resizebox{3.2in}{!}{\includegraphics{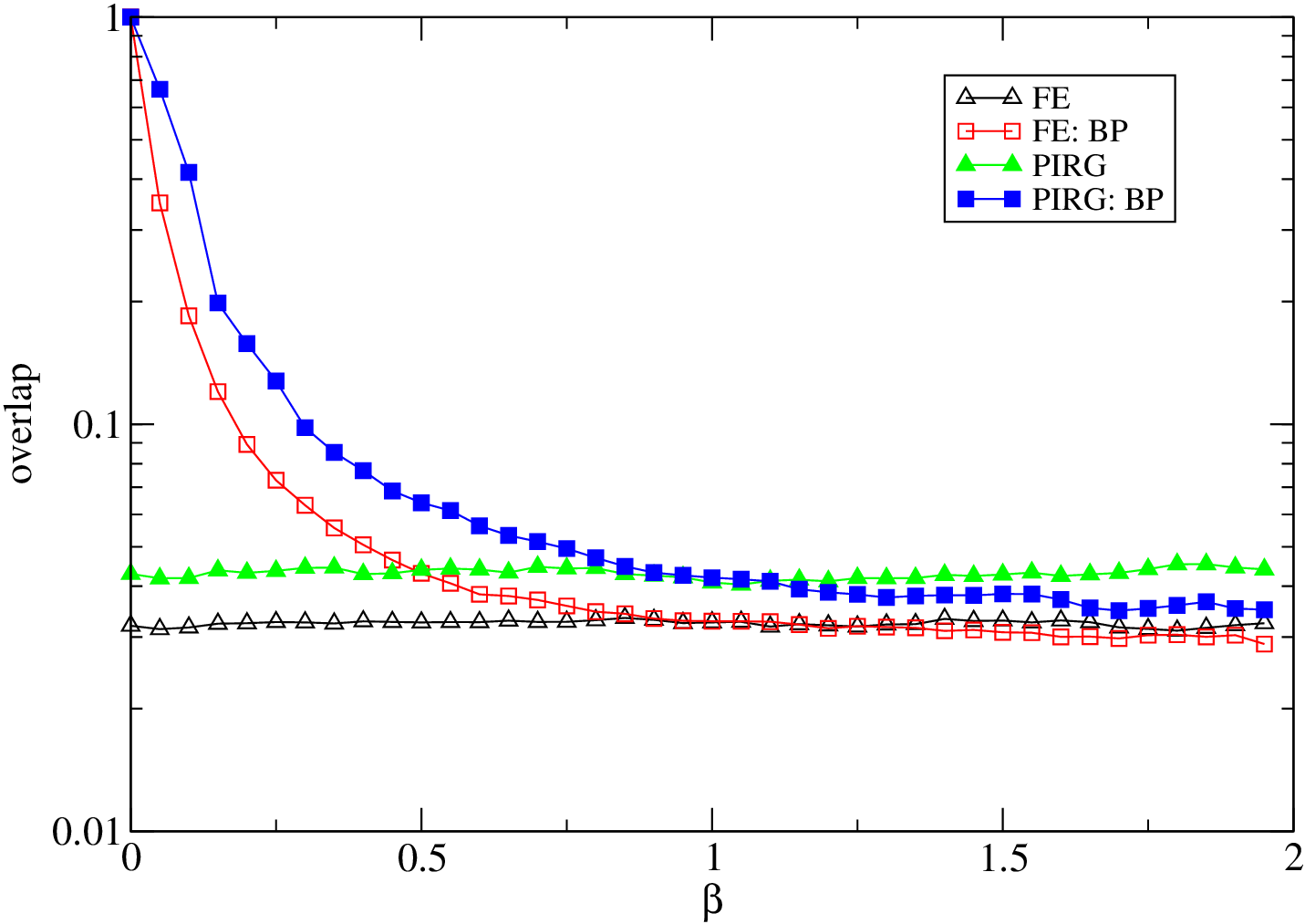}}}
  \caption{(color online) Overlap during back propagation $O_t$
    (see text).  Data are
    for the 4$\times$4 lattice with $N_e$=10, $U$=8, $\Delta\tau=0.05$,
    and 500 walkers. Open (filled) symbols
    used the free-electron
    (PIRG)  wavefunction for $|\Psi_t\rangle$; Triangles (squares)
    show the overlap during forward (reverse) propagation.}
  \label{bp-overlap}
\end{figure}

In the importance sampling scheme used in CPMC, random walkers are weighted by their overlap
with $|\Psi_t\rangle$. The population of random walkers must
periodically be reweighted. We use the ``comb'' reweighting algorithm
detailed in Reference \cite{Gubernatis16a}.
Fig.~\ref{walker-convergence} shows the
$d_{x^2-y^2}$ $\bar{P}$ calculated using BP at $\beta=2$ for the
4$\times$4 lattice at $U=8$ (see Fig.~\ref{4x4u8data}) as a function
of the number of walkers and $\Delta\tau$. In the present paper we
have taken the number of walkers as 100N, or 1600 for the
4$\times$4 lattice. Fig.~\ref{walker-convergence}(a) also shows the
dependence on $\Delta\tau$,
and Fig.~\ref{walker-convergence}(b) shows a typical variation of
 a  BP measurement as a function of $\Delta\tau$.
 In this case, the difference between
$\Delta\tau$=0.01  used in the 4$\times$4 data presented in the main paper
versus a full extrapolation in $\Delta\tau$ is negligible.

During BP, $N_{\rm BM}$ propagators are applied to $|\Psi_t\rangle$
in the reverse order they were used in the original CPMC
random walk. In Fig.~\ref{bp-overlap}, we plot the
  both the summed overlap of the forward-going walkers,
\begin{equation}  
  \sum_i w_i \langle \Psi_t|e^{-\beta\hat{H}(i)}|\phi_i\rangle
  \label{tot_ovlp}
\end{equation}
   and the back-propagated overlap,
\begin{equation}
\sum_i w_i \langle \Psi_t|e^{-\beta\hat{H}(i)}|\Psi_t\rangle
  \label{tot_psit_ovlp}
\end{equation}
as a function of $\beta$ during one BP measurement for the 4$\times$4
system.  In Eq.~\ref{tot_ovlp} and \ref{tot_psit_ovlp}, $\hat{H}(i)$
indicates that the Hubbard-Stratonovich fields are those belonging to
walker $i$.  Fig.~\ref{bp-overlap} shows that $O_t$ becomes
comparable to $\langle \Psi_t|\phi_i\rangle$ for $\beta\sim 1$.
Figs.~\ref{4x4u4data} and \ref{4x4u8data} shows that for $\beta\sim
1$, BP measurements are nearly converged.  Fig.~\ref{bp-overlap} also
demonstrates that with a more accurate PIRG trial wavefunction, the
overlap of random walkers with $|\Psi_t\rangle$ is larger.

\begin{figure}[tb]
  \centerline{\resizebox{3.2in}{!}{\includegraphics{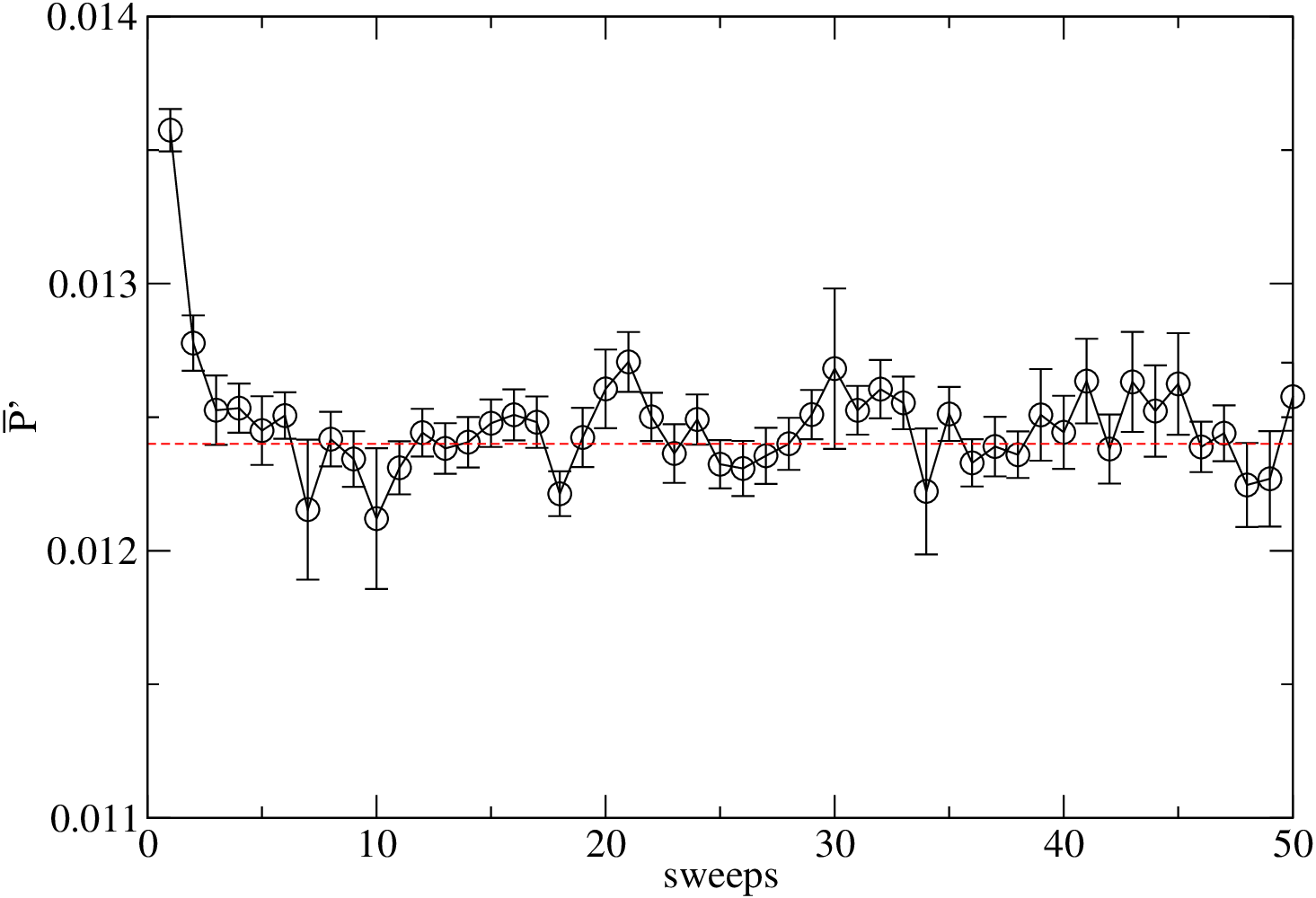}}}
  \caption{(color online) Modified CR $\bar{P}$ estimate
    $\bar{P}'$ (see text) versus the number of Monte Carlo sweeps. Data are
    for the 4$\times$4 lattice with $N_e$=10, $U$=8, $\tau$=0.5, $\beta$=2, and
    1600 walkers. The dashed line is a guide to the eye.}
  \label{cr-convergence}
\end{figure}

The Monte Carlo sampling performed during each CR measurement
introduces more parameters, including the number of measurement
sweeps and the number of equilibration sweeps. Before each CR
measurement, Hubbard-Stratonovich fields ${\bf x}$ are initialized
to the same values used in a BP measurement. These initial fields should
already be more representative of the ground state than
the random initial fields typically used in the projector
quantum Monte Carlo method. However, some number of equilibration
sweeps must still be performed before recording CR data.

Each CR measurement (see Eq.~\ref{cr}) depends on the average sign of
the field configurations along the released path.  In order 
 show how CR measurements depend on the number of sweeps over ${\bf
  x}$, we have omitted this sign factor in Fig.~\ref{cr-convergence}.
Fig.~\ref{cr-convergence} shows the result for the modified $\bar{P}'$
as a function of the number of CR sweeps, averaged over 1600 walkers
and 2200 CR measurements. An initial transient can be seen, lasting
for less than 10 sweeps. In practice, for each CR measurement we
perform 25 equilibration sweeps before taking data, and then take
measurements every 10 sweeps up to a total of 200 sweeps.  This number
of sweeps is enough to well converge the results we present in this
paper, but it is likely that for larger systems fewer CR sweeps can be
performed in order to speed up the method  while only
  slightly increasing statistical errors.

\end{document}